\documentclass[%
 reprint,
 amsmath,amssymb,
 aps,
 superscriptaddress
]{revtex4-2}

\usepackage{graphicx}
\usepackage{dcolumn}
\usepackage{bm}
\usepackage{comment}
\usepackage{physics}
\usepackage{lipsum}
\usepackage{makecell}
\usepackage{color,soul}
\usepackage{pdfpages}

\makeatletter
\AtBeginDocument{\let\LS@rot\@undefined}
\makeatother

\begin{document}

\title{Electrically interfaced Brillouin-active waveguide for multi-domain transduction}

\author{Yishu Zhou}
\affiliation{Department of Applied Physics, Yale University, New Haven, CT, USA}
\author{Freek Ruesink}
\affiliation{Department of Applied Physics, Yale University, New Haven, CT, USA}
\author{Margaret Pavlovich}
\affiliation{Department of Applied Physics, Yale University, New Haven, CT, USA}
\author{Ryan Behunin}
\affiliation{Department of Applied Physics and Materials Science, Northern Arizona University, Flagstaff, AZ, USA}
\author{Haotian Cheng}
\affiliation{Department of Applied Physics, Yale University, New Haven, CT, USA}
\author{Shai Gertler}
\affiliation{Department of Applied Physics, Yale University, New Haven, CT, USA}
\author{Andrew L. Starbuck}
\affiliation{Microsystems Engineering, Science, and Applications, Sandia National Laboratories, Albuquerque, NM, USA}
\author{Andrew J. Leenheer}
\affiliation{Microsystems Engineering, Science, and Applications, Sandia National Laboratories, Albuquerque, NM, USA}
\author{Andrew T. Pomerene}
\affiliation{Microsystems Engineering, Science, and Applications, Sandia National Laboratories, Albuquerque, NM, USA}
\author{Douglas C. Trotter}
\affiliation{Microsystems Engineering, Science, and Applications, Sandia National Laboratories, Albuquerque, NM, USA}
\author{Katherine M. Musick}
\affiliation{Microsystems Engineering, Science, and Applications, Sandia National Laboratories, Albuquerque, NM, USA}
\author{Michael Gehl}
\affiliation{Microsystems Engineering, Science, and Applications, Sandia National Laboratories, Albuquerque, NM, USA}
\author{Ashok Kodigala}
\affiliation{Microsystems Engineering, Science, and Applications, Sandia National Laboratories, Albuquerque, NM, USA}
\author{Matt Eichenfield}
\affiliation{James C. Wyant College of Optical Sciences, University of Arizona, Tucson, AZ, USA}
\author{Anthony L. Lentine}
\affiliation{Microsystems Engineering, Science, and Applications, Sandia National Laboratories, Albuquerque, NM, USA}
\author{Nils Otterstrom}
\affiliation{Microsystems Engineering, Science, and Applications, Sandia National Laboratories, Albuquerque, NM, USA}
\author{Peter Rakich}
\affiliation{Department of Applied Physics, Yale University, New Haven, CT, USA}


\begin{abstract}
    New strategies to convert signals between optical and microwave domains could play a pivotal role in advancing both classical and quantum technologies. 
    Through recent studies, electro-optomechanical systems have been used to implement microwave-to-optical conversion using resonant optical systems, resulting in transduction over limited optical bandwidth.
    Here, we present an optomechanical waveguide system with an integrated piezoelectric transducer that produces electro-optomechanical transduction over a wide optical bandwidth through coupling to a continuum of optical modes. 
    Efficient electromechanical and optomechanical coupling within this system enables bidirectional optical-to-microwave conversion with a quantum efficiency of up to $-$54.16~dB. 
    When electrically driven, this system produces a low voltage acousto-optic phase modulation over a wide ($>$100 nm) wavelength range.
    Through optical-to-microwave conversion, we show that the amplitude-preserving nature inherent to forward Brillouin scattering is intriguing and has the potential to enable new schemes for microwave photonic signal processing.
    We use these properties to demonstrate a multi-channel microwave photonic filter by transmitting an optical signal through a series of electro-optomechanical waveguide segments having distinct resonance frequencies. 
    Building on these demonstrations, such electro-optomechanical systems could bring flexible strategies for modulation, channelization, and spectrum analysis in microwave photonics.
\end{abstract}

\maketitle

\section{\label{sec:intro} Introduction}
\vspace{-3pt}

The ability to exchange information between optical and microwave domains is crucial for classical and quantum signal processing, computing, communication, and networking.
Various methods, such as cavity electro-optics~\cite{rueda2016efficient,hu2021chip}, magneto-optics~\cite{zhu2020waveguide}, and atomic interactions~\cite{kumar2023quantum}, have been developed for interfacing optical and microwave signals.   
Among the different approaches to encode radio-frequency (RF) signals in light, acousto-optical coupling is appealing since GHz frequency elastic waves and optical waves can be confined in the same micro-scale structure to produce efficient coupling.
One versatile  strategy for tailoring acousto-optical coupling is based on stimulated Brillouin scattering, a nonlinear optical process arising from the interaction of light with acoustic waves.
Brillouin-based acousto-optical conversion has been used to demonstrate microwave measurement~\cite{jiang2016wide}, synthesis~\cite{li2013microwave,preussler2013generation}, high-resolution filtering~\cite{stern2014tunable,marpaung2015low,botter2022guided,gertler2022narrowband} and signal processing~\cite{choudhary2017advanced,kim2022chip} within traveling wave systems. 

\vspace{-1pt}

Additionally, phonons can also be electrically driven through piezoelectric materials, providing much higher acoustic intensities and a direct microwave interface.
Recently, significant progress has been made in utilizing cavity electro-optomechanical systems to achieve bidirectional microwave-to-optical quantum transduction~\cite{shao2019microwave,han2020cavity,chu2020perspective,jiang2020efficient,mirhosseini2020superconducting,honl2022microwave,stockill2022ultra}. 
In contrast, traveling wave electro-optomechanical systems, which could offer larger optical bandwidth, increased sensitivity over an extended interaction region, and distinct dynamics, remain unexplored despite their potential for efficient optical sensing and versatile microwave photonic signal processing.

\vspace{-1pt}

Here, we present an electrically interfaced Brillouin-active waveguide that enables efficient and wideband acousto-optic modulation on-chip. 
Significant electromechanical and optomechanical coupling is achieved through acoustic resonance, facilitating bidirectional optical-to-microwave conversion with a quantum efficiency up to $-54.16$ dB.
The absence of a microwave or optical cavity distinguishes this system from conventional triply-resonant electro-optomechanical systems~\cite{mckenna2020cryogenic,yoon2023simultaneous} and ensures the wideband optical operation of our device. 
Under microwave excitation, this system employs an acoustically-driven optical intramodal scattering process to generate a series of cascading sidebands in the frequency domain, leading to phase modulation with a $V_\pi \cdot L$ value of 0.13~V$\cdot$cm over an optical bandwidth exceeding 100 nm.
Operating the system in reverse, the amplitude-preserving nature inherent to this intramodal scattering process enables a scalable channelization method for microwave photonics. 
As a proof of concept, we built a multi-channel microwave photonic filter through an electrically interfaced waveguide with spatially-distributed Brillouin-active segments. 

\vspace{-3pt}

\section{\label{sec:design} Device design}
\vspace{-3pt}

\begin{figure*}
    \includegraphics{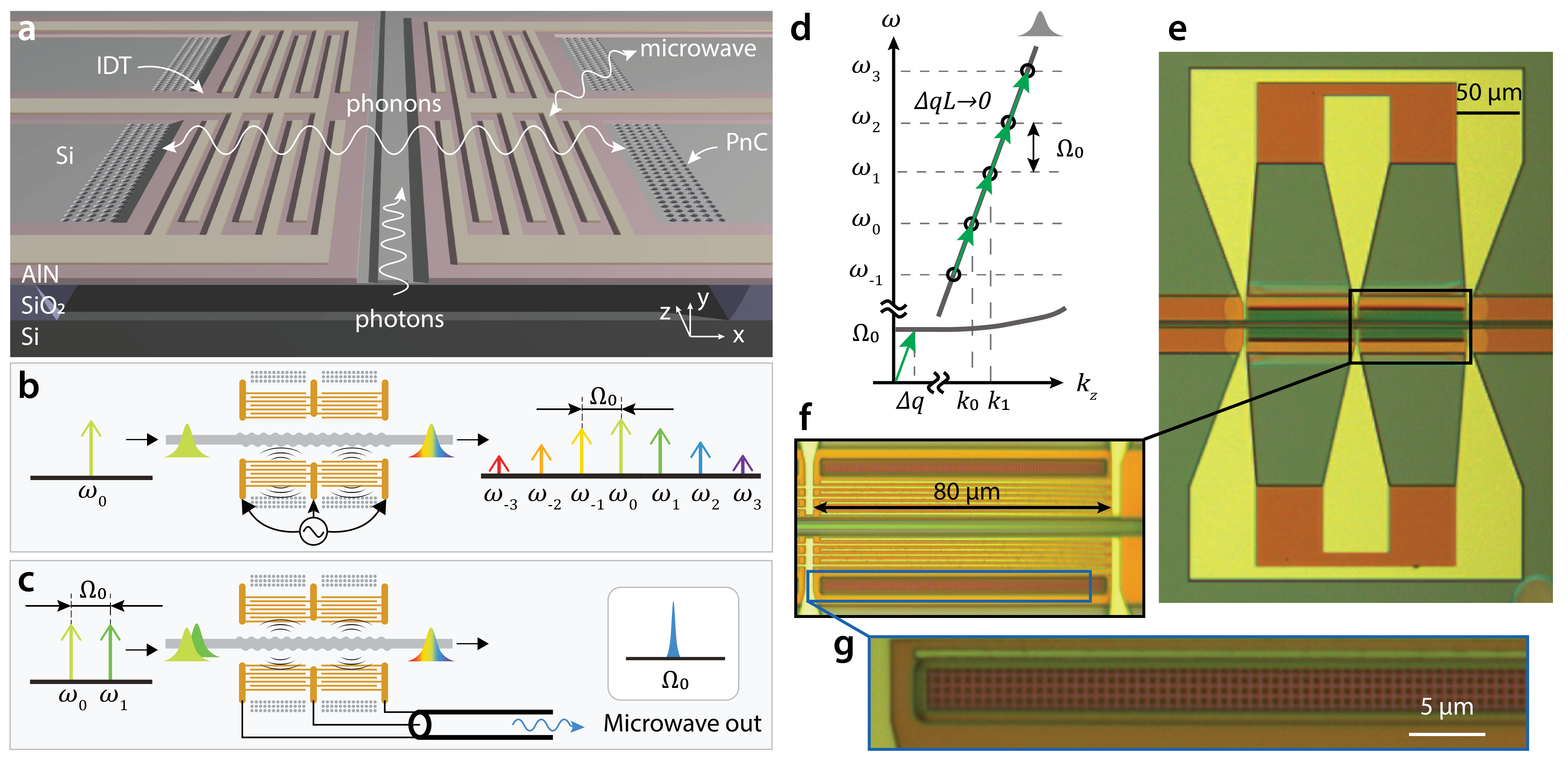}
    \caption{\label{fig1} \textbf{Layout and operation of the electrically interfaced Brillouin-active waveguide. }
    \textbf{a}, A 3D rendering of the device. The photons travel along the single-mode silicon ridge waveguide located in the center of the suspended membrane, while the phonons are confined laterally within the membrane by the phononic crystal structure on either side. Flanking the optical ridge waveguide is a pair of slim interdigitated transducers (IDTs), which enable the device to interface with microwave signals through piezoelectric transduction.  
    \textbf{b, c}, The bidirectionality of electrically interfaced Brillouin scattering process. 
    The phonons can either be electrically driven to produce an acousto-optical comb-like light spectrum (\textbf{b}), or optically excited to transduce the optical signal into a microwave signal via the IDTs (\textbf{c}). 
    Both processes facilitate a phase modulation on the incident light, therefore preserving the optical intensity envelope.  
    \textbf{d}, The dispersion relations of the forward Brillouin scattering process. 
    \textbf{e, f, g}, Optical micrographs of the actual device, which was fabricated using a CMOS foundry process on an AlN-on-SOI platform. }
\end{figure*}

The system under study is composed of a suspended AlN-on-silicon membrane with a silicon ridge waveguide at its center, as illustrated in Fig.~\ref{fig1}a. 
The suspended membrane is enclosed by a phononic crystal, facilitating the confinement for a high-$Q$ phonon mode at frequency $\Omega_0/(2\pi)\sim$ 3.63~GHz. 
Simultaneously, the ridge waveguide provides optical confinement for a low-loss TE-like optical mode in telecom wavelengths. 
A pair of slim IDTs symmetrically flank the optical ridge waveguide, enabling the piezoelectric coupling to acoustic phonons.
The dimensions of the structure (Supplementary Information Sec.~II) are tailored to ensure large acousto-optic and electro-mechanical mode overlaps, thus yielding substantial acousto-optic scattering and electromechanical transduction. 
The entire system is fabricated using the CMOS foundry process reported in~\cite{zhou2022intermodal}. 

The operation scheme of the electrically interfaced Brillouin scattering is shown in Fig.~\ref{fig1}b and c, which demonstrate the bidirectional functionality of this device. 
In Fig.~\ref{fig1}b, a microwave drive at frequency $\Omega_0$ excites elastic waves towards the optical waveguide, with an acoustic wavevector $q = 0$ in the $z$ direction.
These phonons then interact with the light traveling in the optical waveguide via photoelastic scattering. 
In the presence of a strong phonon drive, incident light of frequency $\omega_0$ scatters into several optical sidebands at frequencies $\omega_n=\omega_0 + n\Omega_0$ with $n$ being an integer.
This creates a comb-like optical spectrum. 
Conversely, as depicted in Fig.~\ref{fig1}c, acoustic phonons can be generated through stimulated Brillouin scattering by injecting two optical tones into the waveguide, leading to a microwave signal output through the IDT transduction. 
Both processes induce a phase modulation on the incident light field through variations in the refractive index resulting from phonon-induced density and pressure fluctuations~\cite{wolff2017cascaded}. 
Consequently, this system effectively maintains the amplitude of the incident optical field, ensuring the preservation of any information encoded within the light's intensity envelope. 
This amplitude-preserving characteristic, rooted in intramodal Brillouin scattering, holds significant potential for a range of applications in microwave photonic technologies, which will be further explored in Sec.~\ref{sec:scalability}. 

The phase-matching diagram in Fig.~\ref{fig1}d suggests that to achieve maximum efficiency in Brillouin scattering, we need a wavevector of $q' = k_1 - k_0 = \Omega_0/v_\mathrm{g}$. 
However, there is a small wavevector mismatch, $\Delta q $, between this ideal wavevector and the electrically driven phonon wavevector $q = 0$.
In our device, we intentionally designed a short interaction region to ensure a negligible phase-mismatch $\Delta q L \ll 1$. 
Fig.~\ref{fig1}e, f, and g depict micrographs of a fabricated device, which consists of an acousto-optic interaction region that is 160 \textmu m in length, along with two 2-IDT-units probed in a GSG (Ground/Signal/Ground) configuration. 
Fig.~\ref{fig1}g offers a magnified view of the phononic crystal region, characterized by a cubic lattice of air holes.
In this study, we will demonstrate that our device is capable of efficient acousto-optical phase modulation and enables bidirectional conversion of signals between microwave and optical domains.

\section{\label{sec:characterization} Electromechanical Brillouin scattering}

\begin{figure}[h!]
    \includegraphics{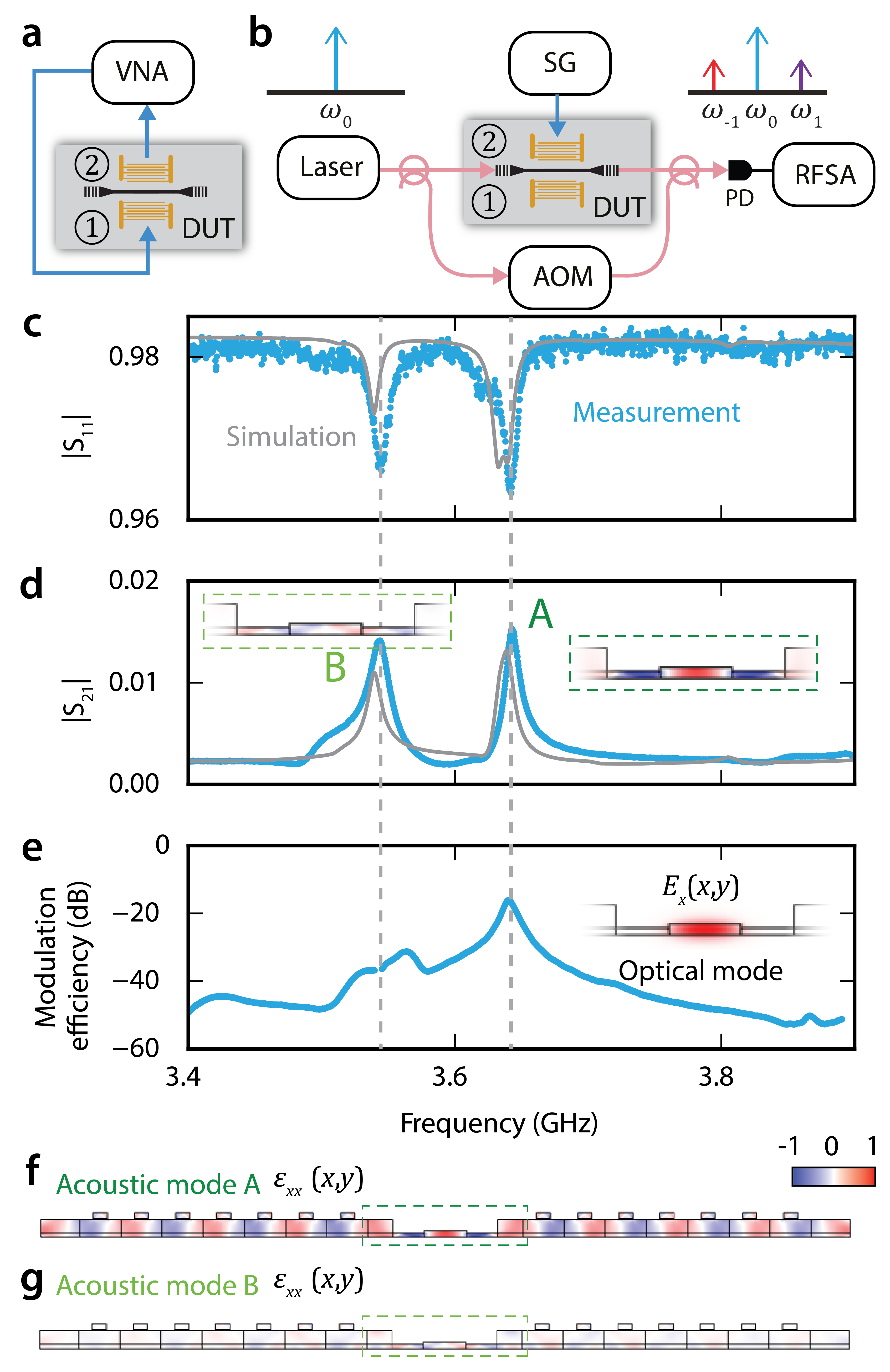}
    \caption{\label{fig2} \textbf{Electromechanical Brillouin scattering of the device. } 
    \textbf{a}, The experimental setup for electrical characterization. VNA: vector network analyzer. 
    \textbf{b}, A heterodyne setup to characterize electromechanical Brillouin scattering. SG: signal generator, AOM: acousto-optic modulator, RFSA: RF spectrum analyzer. 
    \textbf{c, d}, Microwave reflection $|S_{11}|$ (\textbf{c}) and transmission $|S_{21}|$ (\textbf{d}) of the IDTs, obtained from experimental measurements (blue) and finite-element simulations (grey). 
    These traces reveal two distinct electromechanical resonances corresponding to modes A and B, with detailed mode profiles shown as insets in \textbf{d}.
    \textbf{e}, Optical modulation measurement by sweeping the driving RF frequency obtained using the setup shown in \textbf{b}. 
    The modulation efficiency, defined as the conversion efficiency of the 1st sideband $P_\mathrm{out,1}/P_\mathrm{in}$, is proportional to the mode overlaps between the acoustic (\textbf{d} inset) and optical (\textbf{e} inset) modes. 
    As a result, mode A exhibits significantly stronger modulation efficiency compared to mode B.
    \textbf{f, g}, The acoustic profiles ($xx$ component of the strain) of mode A (\textbf{f}) and B (\textbf{g}), given by eigenmode simulations. 
    }
\end{figure}

To understand the behavior of this device, we first analyze its electrically driven operation, i.e., the physical process of electromechanical Brillouin scattering. 
We use a calibrated vector network analyzer (VNA) with two 3-point probes in GSG configuration (100-micron pitch) to measure all four $S$ parameters of a pair of 7-tooth IDTs, as illustrated in Fig.~\ref{fig2}a. 
The measured frequency-dependent microwave reflection of IDT 1 ($|S_{11}|$) and microwave transmission from IDT 1 to IDT 2 ($|S_{21}|$) are shown in blue in Figs.~\ref{fig2}c and d, and align well with the simulated frequency-domain response obtained from COMSOL finite element analysis, shown in gray. 
Our measurements reveal two distinct electromechanical resonances in both the $|S_{11}|$ and $|S_{21}|$ measurements, namely mode A (3.63~GHz) and mode B (3.53~GHz), with the strain profiles shown in the insets of Fig.~\ref{fig2}d.  
Despite a large impedance mismatch to the driving circuit caused by the compact IDT design, the IDT demonstrates appreciable electro-acoustic transduction efficiency, which is improved $\sim4\times$ relative to a reference IDT without external acoustic resonances (Supplementary Information Sec.~III).

Fig.~\ref{fig2}f~(g) shows the full strain profile of mode A~(B), as predicted by eigenmode finite-element simulations. 
In order to obtain a large acousto-optic coupling rate ($g$), it is important that the symmetry of the strain profile ($\epsilon_\mathrm{xx}$) aligns well with the optical mode profile ($E_\mathrm{x}$) to create good mode overlap, since $g \propto \int \epsilon_\mathrm{xx} E_\mathrm{x} \, dV$.  
The different spatial profiles of $\epsilon_\mathrm{xx}$ in modes A and B dictate their varying Brillouin scattering response.
Comparing the zoomed-in acoustic profiles in the insets of Fig.~\ref{fig2}d, mode A exhibits a pronounced and symmetrical $\epsilon_\mathrm{xx}$ strain component, whereas mode B shows a weaker and anti-symmetrical one.
Therefore, mode A allows for a much better mode overlap with the optical mode profile $E_x$ (Fig.~\ref{fig2}e inset), resulting in stronger acousto-optic scattering. 

We verify our simulation results by measuring the corresponding electromechanical Brillouin scattering when driving IDT 2 with a microwave power of 8.92~dBm using the heterodyne measurement setup (Fig.~\ref{fig2}b). 
Here the modulation efficiency is defined as $P_\mathrm{out,1}/P_\mathrm{in}$, where $P_\mathrm{out,1}$ is the optical power in the 1st sideband at the output of the device, and $P_\mathrm{in}$ the optical power that is injected into the device (approximately 1~mW in this experiment). 
As illustrated in Fig.~\ref{fig2}e, for an incident microwave power of 8.92~dBm, the optical scattering efficiency reaches its maximum of $-16.32$~dB at 3.63~GHz (mode A), indicating strong electromechanical Brillouin scattering. 
This sets the foundation for the acousto-optic phase modulator that we will analyze in detail in Sec.~\ref{sec:phaseMod}.
Conversely, due to opposing symmetries of the optical and acoustic modes, mode B exhibits a dip in the measured spectrum, leading to a significantly lower scattering efficiency than mode A, with a decrease of over 20~dB. 

\section{\label{sec:phaseMod} Optical phase modulation}

\begin{figure}
    \includegraphics{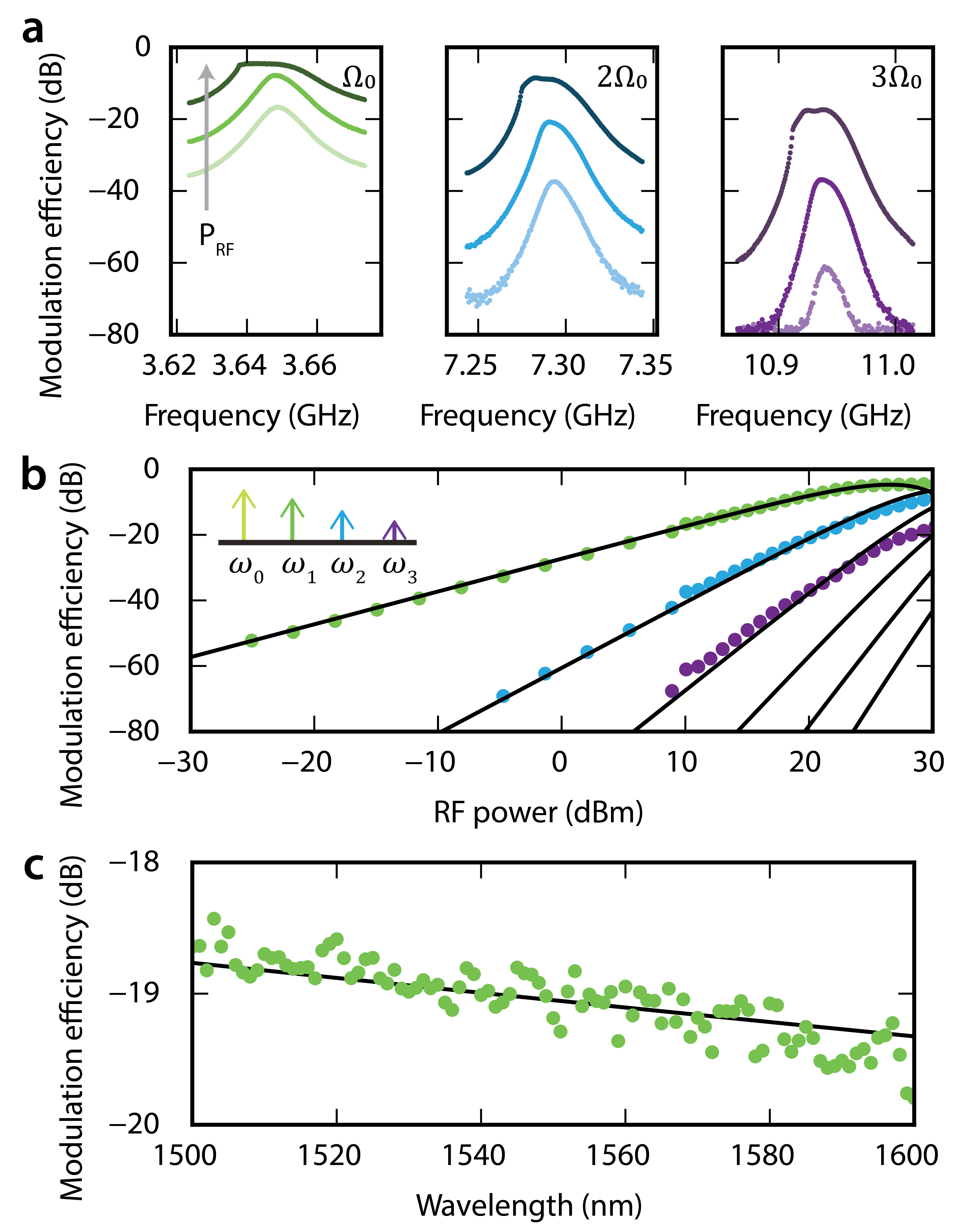}
    \caption{\label{fig:PM} \textbf{Optical phase modulation. }
    \textbf{a}, Modulation efficiency of the first three sidebands.
    The modulation spectrum of each sideband was obtained at three different RF power levels $-$10.06 dBm, 20.12 dBm, and 30.24 dBm---indicated by lighter to darker colors, respectively. 
    \textbf{b}, RF Power sweep of modulation efficiency at resonance, which closely follows the theoretical predictions of Bessel functions.
    We obtain a $V_\pi = 8.13$ V and $V_\pi L = 0.13 $ V$\cdot$cm. 
    \textbf{c.} Modulation of the first sideband shows minimal variation over a wavelength range of 100 nm. } 
\end{figure}

The strong electrically driven acousto-optic intramodal scattering induces an efficient phase modulation across the entire photonic band. 
Assuming a monochromatic incident optical field with frequency $\omega_0$ and amplitude $a_0$, the phase-modulated optical output at $z=L$ is given as (Supplementary Information Sec.~I)
\begin{equation}
    \tilde{a}_{\mathrm{out}} (L, t) = a_0 \exp \left[ - i \beta \cos \left( \Omega_0 t \right)  \right], 
    \label{eq:PM}
\end{equation}
where the rotating frame $\tilde{a}_{\mathrm{out}} (L, t) = a_{\mathrm{out}} (L, t) e^{i \omega_0 t - i k_0 L}$ is introduced. 
The modulation depth, $\beta = 2 g \left|b\right| L/v_\mathrm{g} $, where $v_\mathrm{g}$ is the group velocity of the optical mode, is determined by the phonon amplitude, which is controlled using the microwave drive. 
Under the RF drive with power $P_{\mathrm{RF}}$ and frequency $\Omega$, the excited phonon amplitude is given by $b(\Omega) \propto \chi_\mathrm{b}(\Omega) \sqrt{\eta_\mathrm{em} P_{\mathrm{RF}}}$, where $\eta_\mathrm{em}$ is coupling factor of the electromechanical system. 
In our experiment, $\eta_\mathrm{em}$ is characterized by the microwave transmission $|S_{21}|$ from one IDT to the other, as measured in Fig.~\ref{fig2}d. 
The phonon susceptibility, $\chi_\mathrm{b}(\Omega)$, is defined as $\chi_\mathrm{b}^{-1} (\Omega) = -i(\Omega-\Omega_0)+\Gamma_0/2$, and $\Gamma_0$ is the phonon linewidth.

The phase modulated optical output can be described as a series of optical tones with frequency $\omega_n$, with amplitudes given by the Bessel decomposition, 
\begin{equation}
    \tilde{a}_n (L, t) = i^n a_0 J_n \left( - \beta \right) e^{-i n \Omega_0 t}.
    \label{eq:a_n}
\end{equation}
The modulation efficiency $|a_n|^2/|a_0|^2$ of the first three sidebands is presented in Fig.~\ref{fig:PM}a, which is measured using the same 7-tooth device and experimental setup as in Fig.~\ref{fig2}b. 
The modulation efficiency produced over a range of drive frequencies is shown for three different RF powers, 10.06 dBm, 20.12 dBm, and 30.24 dBm, which are depicted from lighter to darker colors.
At low RF power, each optical tone exhibits a Lorentzian-like frequency-dependent modulation amplitude due to the presence of the phonon resonance. 
As the RF power is increased, the phase modulation gradually gets saturated, and the measured acoustic waveshape gets distorted and thermally shifted from the standard Lorentzian shape (Supplementary Information Sec.~IV). 
The modulation efficiencies at 3.63~GHz as a function of RF power are presented in Fig.~\ref{fig:PM}b, which agree well to the theoretical predictions from Bessel functions in black. 
Overall, our compact optically-nonresonant system achieves a significant effective electro-optic coefficient with half-wave voltage $V_\pi = 8.13\pm0.48$ V and $V_\pi L = 0.1301\pm0.0077$ V$\cdot$cm. 
Note that the higher-order frequency tones ($n>3$) are also present in this system as shown by the black curves in Fig.~\ref{fig:PM}b. 
These tones are not measured because they fall outside the frequency range of our fast photodetector (Nortel PP-10G).

The absence of an optical cavity also enables wideband operation of our acousto-optic phase modulator, which is characterized in Fig.~\ref{fig:PM}c. 
We characterize the first sideband modulation efficiency under the RF drive of 8.92 dBm, which shows less than 1 dB variation over 100 nm wavelength range. 
The measured efficiency agrees well with a $1/\lambda^2$ trend as denoted by the black line, which is caused by the acousto-optic overlap change at different optical wavelengths~\cite{kittlaus2021electrically}.

\section{\label{sec:OE} Bidirectional conversion}
\begin{figure}
    \includegraphics{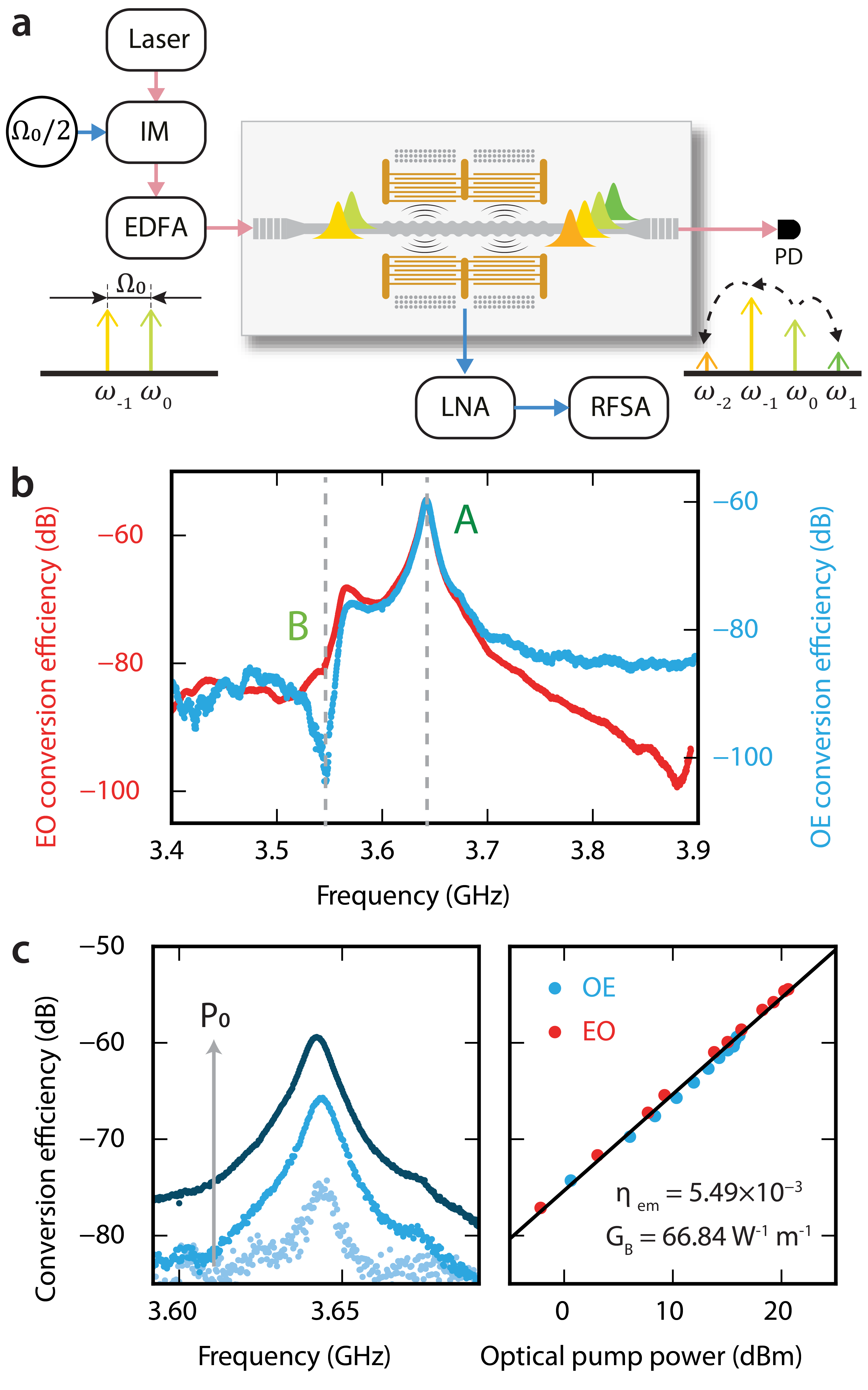}
    \caption{\label{fig:OE} \textbf{Bidirectional conversion. }
    \textbf{a}, Experimental setup for characterizing the optical-to-microwave conversion using a dual-pump configuration.
    LNA: low noise amplifier. 
    \textbf{b},  The quantum efficiency of optical-to-microwave conversion characterized at an optical pump power of 16.12 dBm (blue) and microwave-to-optical conversion characterized at an optical pump power of 20.61 dBm (red), with the peak transduction corresponding to mode A. 
    \textbf{c}, The peak of the optical-to-microwave conversion spectra at three different optical powers (left), along with an optical pump power sweep for both the optical-to-microwave (blue) and microwave-to-optical (red) conversions (right). 
    The linear fitting (black) suggests a decent value of Brillouin gain $G_\mathrm{B}$. }
\end{figure}

In addition to electrically driven electromechanical scattering, appreciable electro-optomechanical coupling permits the system to operate in reverse, allowing for optically driven electromechanical transduction.
Through optical-to-microwave transduction, phonons generated through a stimulated Brillouin scattering process are converted to microwave photons by the IDT.
In contrast to resonator-based systems, optical-to-microwave conversion becomes possible over wide optical bandwidths through coupling to a continuum of optical modes. 

Since this is a continuous optomechanical system, conventional cavity-based treatments of electro-optomechanical conversion are not applicable. Instead, it is necessary to formulate couplings and cooperativities based on coupling to continuous bands of states. Through prior studies of optomechanical cooling in continuous systems, it has been shown that the effective opto-mechanical cooperativity is given by $C_\mathrm{om} \cong |g a_0|^2L / (v_\mathrm{g} \Gamma_0) = G_\mathrm{B}P_0L/4$, where $G_\mathrm{B}$, $P_0$, and $L$ are the Brillouin gain coefficient, the optical pump power, and the Brillouin interaction length, respectively~\cite{otterstrom2018optomechanical}. 

Building on continuum treatments, we derive an expression for quantum efficiency of electro-optomechanical conversion, which can be expressed as 
\begin{equation}
    \eta = \frac{1}{2} \eta_{\mathrm{em}} G_\mathrm{B} P_0 L. 
    \label{eq:eta}
\end{equation}
See Supplementary Information Sec.~I for details.
Interestingly, this conversion efficiency is reminiscent of traditional cavity electro-optomechanical systems, $\eta = \eta_\mathrm{em} \eta_\mathrm{om} \frac{4 C_\mathrm{om}}{1+C_\mathrm{om}^2}$, in the limit when $C_\mathrm{om}\ll 1$~\cite{han2021microwave}. 
Since our optical waveguide has negligible losses over the short ($160$ micron) interaction length, all photons directly transit through the system, leading to a unity external coupling factor for the optomechanical system ($\eta_\mathrm{om} = 1$).
Moreover, the factor of $\frac{1}{2}$ in Eq.~(\ref{eq:eta}) arises from the presence of two sets of IDTs; microwave energy is emitted by each IDT into two microwave output ports, and we only detect the microwave energy from one of these ports.

We characterized the optical-to-microwave conversion via a dual-pump stimulated Brillouin process, using the setup depicted in Fig.~\ref{fig:OE}a.
Two drive fields, at equal intensities and frequencies of $\omega_{-1}$ and $\omega_0$, are produced via a carrier-suppressed intensity modulator, which is driven at a frequency of $\Omega_0/2$.
As these fields propagate through the optical linear waveguide, they couple with each other through parametrically generated acoustic phonons with a Brillouin frequency of $\Omega_0$. 
These acoustic phonons extend throughout the membrane structure and traverse the IDTs, with the piezoelectric substrate converting the mechanical energy of the phonons into an electrical signal, producing a transduced microwave response at the output of the IDTs. 
The resulting microwave signal is then amplified via a low noise amplifier (LNA) and read out by a spectrum analyzer.

Fig.~\ref{fig:OE}b shows the frequency-dependent measurement of the conversion efficiency of another 5-tooth IDT device. 
The red curve represents the microwave-to-optical conversion efficiency (EO), while the blue curve represents the optical-to-microwave conversion efficiency (OE). 
Both curves exhibit peak conversion at 3.63 GHz, corresponding to acoustic mode A as shown in Fig.~\ref{fig2}f. 
We measured the maximum OE efficiency to be $-59.24\pm$0.24 dB at an optical pump power of 16.12 dBm, and the maximum EO efficiency to be $-54.16\pm$0.24 dB at an optical pump power of 20.61 dBm, highlighting the efficient operation of our device.
We note that the OE measurement shows a significant cancellation dip at mode B, which corresponds well to the nullified acousto-optic coupling rate as discussed in Sec.~\ref{sec:characterization}. 

To further understand the system properties, we performed an optical pump power sweep as shown in Fig.~\ref{fig:OE}c. 
The left panel displays the zoomed-in frequency sweeps of the optical-to-microwave efficiency at different optical power levels, namely 1.16~mW, 10.87~mW, and 48.92~mW. 
The negligible change in the Lorentzian shape of the spectra suggests that the system experiences minimal thermo-optic heating, demonstrating the advantage of employing a linear waveguide system.
The right panel exhibits the well-aligned quantum effciency of both the optical-to-microwave (blue) and microwave-to-optical (red) conversions as a function of optical pump power. 
Considering the measured electromechanical coupling factor of $\eta_\mathrm{em} = 5.49\times10^{-3}$, we determine a Brillouin gain of $G_\mathrm{B} = 66.84 \pm 3.63$ W$^{-1} \cdot$ m$^{-1}$ through linear fitting (black line).
Notably, it should be acknowledged that this value is approximately four times smaller than our COMSOL simulations. 
This discrepancy may arise from fabrication imperfections, suboptimal acousto-optical overlaps, and reduced photoelastic constants of strained silicon, which indicates significant room for improvement in our device (Supplementary Information Sec.~II).

\section{\label{sec:scalability} Microwave photonic channelization }
\begin{figure*}
    \includegraphics{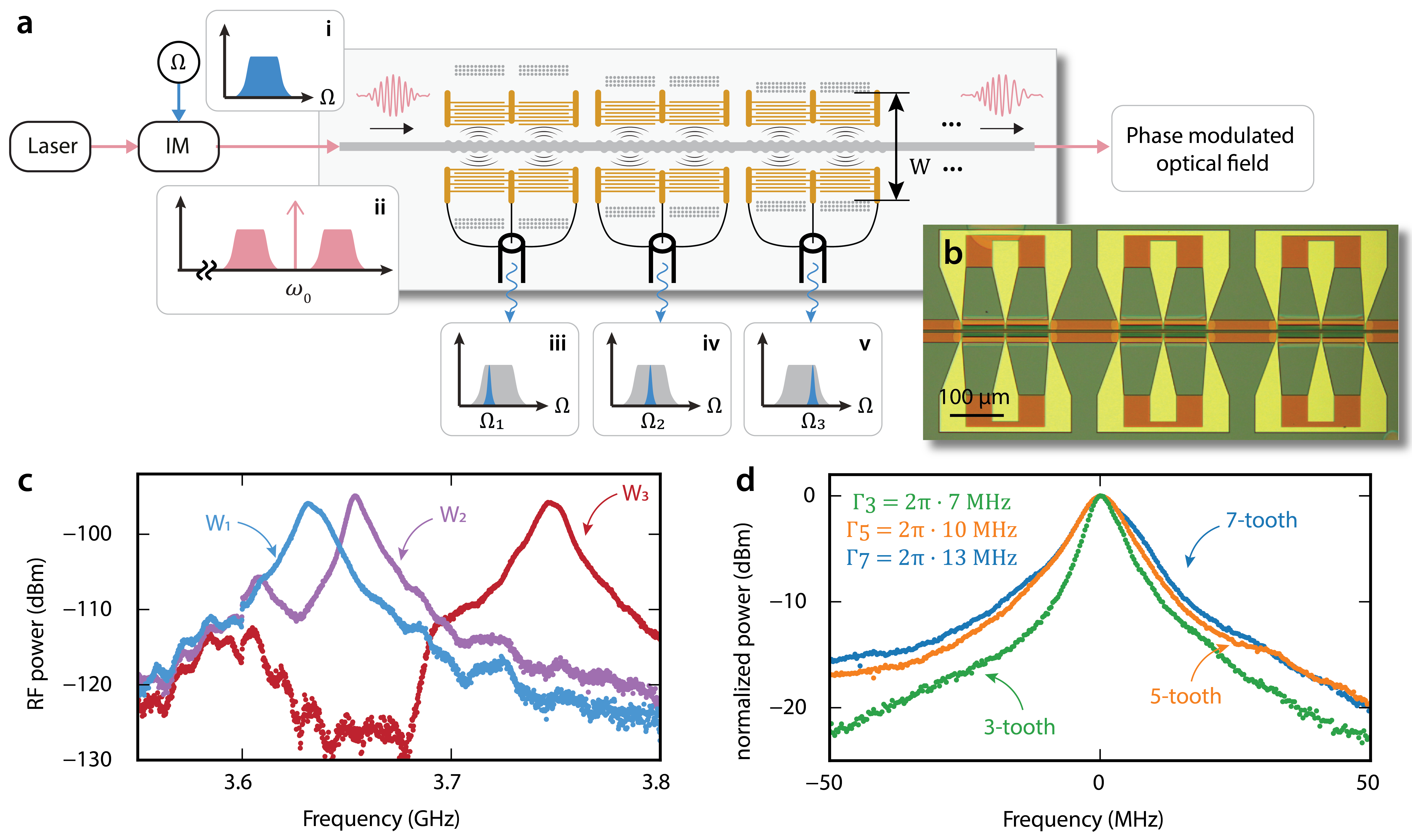}
    \caption{\label{fig:scalability} \textbf{Microwave photonic channelization. }
    \textbf{a}, Operating principle of a multi-segment electrically interfaced Brillouin-active waveguide for microwave photonic channelization. 
    Intensity-modulated light (ii) carrying a wideband RF signal (i) is injected into the system, which excites localized phonons at each electro-optomechanical region. 
    Each phonon generates a microwave signal output with a designed center frequency and targeted bandwidth (iii–v).
    Despite undergoing several phase modulations, the output optical field preserves the intensity envelope, ensuring that the RF information encoded in the optical signal remains unaltered.
    \textbf{b}, Optical micrograph of the multi-channel device. 
    \textbf{c.} Tunability of the microwave filter center frequency using the membrane width ($W$) of 16.15~\textmu m, 15.43~\textmu m, and 14.70~\textmu m. 
    \textbf{d.} Tunability of the microwave filter bandwidth using the number of IDT teeth. }
\end{figure*}

This form of optical-to-microwave conversion is an intriguing alternative to fast photodetectors as a means of extracting the microwave signal encoded on light.
Since the phonons produced intramodal Brillouin scattering process can only produce phase modulation of the light traversing the system, information encoded on the intensity of the light field is preserved (see Supplementary Information Sec.~I). 
As a consequence, information encoded via intensity modulation is preserved even when traversing an array of such Brillouin-active waveguide segments having distinct Brillouin frequencies, as illustrated in Fig.~\ref{fig:scalability}. 
These dynamics permit us to construct a microwave photonic channelizer by cascading Brillouin-active waveguide segments while reading out the transduced microwave signals at different frequencies. 

The operation scheme of the multi-channel device is diagrammed in Fig.~\ref{fig:scalability}a. 
A sequence of electro-optomechanical regions is arranged along one linear waveguide, each supporting a distinct acoustic eigenmode with resonant frequency $\Omega_n$ and linewidth $\Gamma_n$. 
These properties can be adjusted through the geometric features of each region, such as the membrane width $W$ and the number of IDT teeth, providing flexibility and tunability in the device design. 
A wideband RF signal (Fig.~\ref{fig:scalability}a.i) is encoded in the optical carrier (frequency $\omega_0$) through intensity modulation, generating a light field with optical spectrum depicted in Fig.~\ref{fig:scalability}a.ii. 
The modulated light is then injected into the waveguide and excites localized phonons at a different frequency in each section, resulting in precise narrowband microwave readouts in each channel (Fig.~\ref{fig:scalability}a.iii to v). 
These optically driven phonons will each generate a pure phase modulation on the incident optical signal, 
resulting in an output field given as $a_\mathrm{out} = a_\mathrm{in} \exp \left[ - i \sum_n \beta_n \cos \left( \Omega_n t \right) \right]$ (Supplementary Information Sec.~I). 
Here $\beta_n$ is the modulation depth of the $n$th section, determined by the acoustic properties of the $n$th membrane. 
This implies that the original signal, the wideband RF signal injected into the intensity modulator, remains unchanged regardless of the presence of various Brillouin nonlinear interactions of different strengths and frequencies. 

To experimentally validate the channelization concept, we measured the device shown in Fig.~\ref{fig:scalability}b, where each active segment is 160 \textmu m in length and separated by 130 \textmu m.
We controlled the center frequency of each filtering channel by adjusting the membrane width, as shown in Fig.~\ref{fig:scalability}c. 
Specifically, the transduced RF signal is measured from three segments with membrane widths of 16.15~\textmu m, 15.43~\textmu m, and 14.70~\textmu m, corresponding to center frequencies of 3.746~GHz, 3.654~GHz, and 3.631~GHz, respectively. 
These measurements were taken using 3-tooth IDT devices with the parameters listed in Supplementary Information Sec.~II, providing similar transduction efficiency across all channels. 

In our device, the major source of acoustic loss is metal scattering, making the number of IDT teeth a direct candidate for tuning the device's bandwidth. 
Fig.~\ref{fig:scalability}d illustrates the characterization of active segments with 3, 5, and 7 IDT teeth, indicating a range of filtering bandwidths from 7 MHz to 13 MHz and demonstrating the design flexibility of the proposed method.

In summary, we successfully developed a microwave channelizer featuring tunable center frequency and filtering bandwidth for each channel. 
The feasibility of our proposal is grounded in the low-loss optical waveguide and compact form factor of each active segment. 
In our current setup, the optical propagation loss is negligible, making it possible to cascade 100 channels while maintaining minimal total propagation loss and acceptable phase-mismatch loss (Supplementary Information Sec.~II).
As we consider larger systems, it is also interesting to note that the primary noise source, produced by spontaneous Brillouin scattering, also only impacts the phase information encoded on the optical wave. 
Hence, Brillouin scattering produced by this intra-modal scattering process does not degrade the information encoded on the intensity of the light field.

\section{\label{sec:conc} Discussion and conclusion}

We have used an electrically interfaced Brillouin-active waveguide to demonstrate a new strategy for bi-directional transduction of signals between optical and microwave domains. 
The unique dynamics of intra-modal Brillouin scattering were used to create a channelizer within a tapped signal processing architecture by cascading an array such waveguide-based transducers. 
While further work is required to reach the high transduction efficiencies ($\eta \sim 0.1$) and small half wave voltages ($V_\pi \sim 19$~mV) demonstrated using resonator-based systems~\cite{zhao2022enabling}, the target application for these approaches are likely to be very different. 
For example, the high power-handling capability ($>100$ mW) and broad bandwidths of optical transduction ($\sim 100$ nm) produced by this waveguide system open the door to new schemes for high dynamic range microwave photonic signal processing. 

To further enhance the electromechanical transduction efficiency $\eta_\mathrm{em}$ of this system we can greatly improve electro-mechanical coupling to the Brillouin-active phonon mode. 
The microwave extinction $\eta_\mathrm{e} = 1 - \left| S_{11} (\Omega)\right|^2$ of the electromechanical transduder, which characterizes the efficiency with which microwaves are converted to phonons, is currently $\sim$2\% due to the large impedance mismatch between the compact IDT design and the standard 50 $\Omega$ input impedance. 
Increasing the system's impedance or changing to a stronger piezoelectric material such as LiNbO$_3$ or AlScN could enhance the efficiency of this conversion up to 50$\times$. 
Additionally, we must increase the conversion efficiency of the microwave energy into the target acoustic mode, given by $\eta_\mathrm{m} = \eta_\mathrm{em}/\eta_\mathrm{e} \sim $ 0.27 to 0.41. 
Optimizing the IDT and acoustic membrane design can improve this efficiency at least $3 \times$ (Supplementary Information Sec.~II). 

Extending the device length is another way to improve the performance of our system, since the phase modulation $V_\pi$ is proportional to $\sqrt{L}^{-1}$ and the transduction efficiency $\eta$ is proportional to $L$. 
For the linear waveguide, the interaction length could be increased by at least a factor of 10 before the phase mismatch becomes significant. 
Moreover, the current measured Brillouin gain $G_\mathrm{B}$ is at least a factor of 4 smaller than the simulated values, indicating significant potential for improving the design to match the best electromechanical resonance and the acoustic modes with optimum acousto-optical overlaps in future devices.
By combining these improvements, we can pave the way for a device with $\eta >-20$ dB and $V_\pi \sim 104$ mV. 

With these performance enhancements, such could enable wideband acousto-optical comb generation~\cite{diddams2020optical}, offering high power handling capacity, MHz range frequency tunability, and CMOS compatibility.
Alternatively, such systems could become the basis for a scalable and versatile platform for optical-to-microwave conversion that preserves information encoded on the intensity of the light field.
In addition to this intra-modal scattering process, device strategies that make use inter-modal Brillouin scattering could open the door to an array of applications including quantum and classical transduction and communication, microwave photonic spectral analysis, and optical sensing.

\section*{Acknowledgements}

We thank Yanni Dahmani, Taekwan Yoon, Betul Sen, and Naijun Jin for useful technical discussions involving Brillouin interactions and IDT designs. 

\textbf{Funding.}
This research was developed with funding from the Defense Advanced Research Projects Agency (DARPA). 
The views, opinions and/or findings expressed are those of the author and should not be interpreted as representing the official views or policies of the Department of Defense or the U.S. Government.
Distribution Statement A - Approved for Public Release, Distribution Unlimited. 
This material is based upon work supported by the Laboratory Directed Research and Development program at Sandia National Laboratories. 
Sandia National Laboratories is a multi-program laboratory managed and operated by National Technology and Engineering Solutions of Sandia, LLC., a wholly owned subsidiary of Honeywell International, Inc., for the U.S. Department of Energy's National Nuclear Security Administration under contract DE-NA-0003525. 
This paper describes objective technical results and analysis. 
The views, opinions, and/or findings expressed are those of the authors and should not be interpreted as representing the official views or policies of the U.S. Department of Energy, U.S. Department of Defense, or the U.S. Government.

\bibliography{bibfile}

\onecolumngrid


\section*{Supplementary material}

# Supplement for "Electrically interfaced Brillouin-active waveguide and multi-domain transduction"

## I. DYNAMICS OF ELECTRICALLY INTERFACED BRILLOUIN-ACTIVE WAVEGUIDE

In this section, we describe a simple model that captures the mode evolution dynamics of our electrically interfaced Brillouin-active waveguide. We will start with the Hamiltonian $\mathcal{H} = \mathcal{H}_{\text{opt}} + \mathcal{H}_{\text{ph}} + \mathcal{H}_{\text{int}}$ for forward Brillouin scattering process [1, 2], given as

$$\mathcal{H}_{\text{opt}} = \sum_{\gamma} \int dk\hbar\omega(k)a_{\Gamma_0 k}^{\dagger}a_{\gamma k}, \tag{S1}$$

$$\mathcal{H}_{\text{ph}} = \sum_{\alpha} \int dq\hbar\Omega(q)b_{\alpha q}^{\dagger}b_{\alpha q}, \tag{S2}$$

$$\mathcal{H}_{\text{int}} = \sum_{\alpha,\gamma,\gamma'} \int \frac{dkdk'dq}{(2\pi)^{3/2}} a_{\Gamma_0 k}^{\dagger}a_{\gamma' k'}b_{\alpha q} \int dz\hbar g(\Gamma_0 k; \gamma' k'; \alpha q)e^{i(k'-k+q)z} + \text{H.c.}. \tag{S3}$$

Here, $b_{\alpha q}$ and $a_{\Gamma_0 k}$ represent the acoustic-mode amplitude operator and the optical amplitude operator. $\gamma, \gamma'$, and $\alpha$ represent the optical and phonon mode indices, with wave vectors $k, k'$, and $q$, respectively. In our intramodal Brillouin scattering system, we consider a single optical TE0 mode and a single acoustic membrane Lamb-like mode, which allows for a simplification of the mode indices summation. Additionally, the coupling rate $g(\Gamma_0 k; \gamma' k'; \alpha q)$ can be regarded as a constant within a reasonable wavelength range in our low optical dispersion system, such that $g = g(\Gamma_0 k; \gamma' k'; \alpha q)$. Consequently, we can simplify the Hamiltonian as

$$\mathcal{H} = \int dk\hbar\omega(k)a_k^{\dagger}a_k + \int dq\hbar\Omega(q)b_q^{\dagger}b_q + \hbar g \int \frac{dkdq}{\sqrt{2\pi}} a_k^{\dagger}a_{k-q}\left(b_q + b_{-q}^{\dagger}\right). \tag{S4}$$

To describe the dynamics of a wideband optical envelope that might span several Stokes shifts, we transform to spatial domain by introducing envelope operators $a(z,t)$ and $b(z,t)$

$$a(z,t) = \int \frac{dk}{\sqrt{2\pi}}a_k e^{ikz}, \quad b(z,t) = \int \frac{dq}{\sqrt{2\pi}}b_q e^{iqz}. \tag{S5}$$

As a result, the optical (acoustic) power flow in the waveguide is written as $P^{\text{opt}} = \hbar\omega_0 v_{\text{g}}a^{\dagger}(z,t)a(z,t)$ ($P^{\text{ph}} = \hbar\Omega_0 v_{\text{g,b}}b^{\dagger}(z,t)b(z,t)$), where $v_{\text{g}}$ ($v_{\text{g,b}}$) is the group velocity of the photons (phonons).

In addition, we introduce the propagation operators $\mathcal{L}_a$ and $\mathcal{L}_b$ to describe the spatial domain Hamiltonian [3],

$$\mathcal{L}_a = \hbar\omega_0 + \hbar\sum_{n=1}^{\infty}\frac{(-i)^n}{n!}\frac{\partial^n\omega}{\partial k^n}\Big|_{k_0}(\partial_z - ik_0)^n, \quad \mathcal{L}_b = \hbar\Omega_0 + \hbar\sum_{n=1}^{\infty}\frac{(-i)^n}{n!}\frac{\partial^n\Omega}{\partial q^n}\Big|_q(\partial_z - iq)^n, \tag{S6}$$

where we assume that all optical carriers have the same frequency $\omega_0$ because the modulation signal, even if it can be wideband, should have a frequency that is at least $10^{-7}$ smaller than the optical frequency. Therefore, the Hamiltonian of the Brillouin scattering process described using the spatial operators is

$$\mathcal{H} = \int dz a^{\dagger}(z,t)\mathcal{L}_a a(z,t) + \int dz b^{\dagger}(z,t)\mathcal{L}_b b(z,t) + g\int dz a^{\dagger}(z,t)a(z,t)\left[b(z,t) + b^{\dagger}(z,t)\right]. \tag{S7}$$

This Hamiltonian is reminiscent of cavity optomechanical systems operating in the zero-detuning regime, where both the blue sideband and the red sideband are present in the system [4]. In this system, the mechanical displacement induces a phase shift on the light field, while leaving the optical amplitude unchanged. Similarly, in our waveguide, the intensity envelope of the photons, with its change described by

$$(\partial_t - v_{\text{g}}\partial_z)\left(a^{\dagger}a\right) = 0, \tag{S8}$$

also remains time-independent during the scattering process. This similarity draws an analogy between our amplitude-preserving method and non-demolition measurements in cavity-optomechanics experiments.

The same conclusion can be verified by directly solving the equations of motion for photons and phonons, which are given as

$$(\partial_t + v_{\rm g}\partial_z)\,\tilde{a} = -ig\left(\tilde{b} + \tilde{b}^\dagger\right)\tilde{a}, \tag{S9}$$

$$\left[\partial_t - i(\Omega - \Omega_0) + \frac{\Gamma}{2}\right]\tilde{b} = -ig\tilde{a}^\dagger\tilde{a} + \sqrt{\xi}B_{\rm in}, \tag{S10}$$

where the rotating frame operators are introduced as $\tilde{a} = a\exp(i\omega_0 t - ik_0 z)$ and $\tilde{b} = b\exp(i\Omega t - iqz)$ to eliminate both the temporal and spatial carriers. $\Gamma$ and $\Omega_0$ respectively represent the temporal power decay rate and the eigenfrequency of the phonon field. To simplify these assumptions, several assumptions are introduced: (1) The optical propagation operator is reduced to the first order due to the linear optical dispersion of our waveguide; (2) The acoustic propagation operator is simplified to the zeroth order owing to the negligible acoustic group velocity $v_{\rm g,b}$; (3) The optical propagation loss is disregarded within the system, considering the excellent low-loss characteristic of our waveguide and the short interaction region in our design. The right-hand-side of eq. (S10) contains two terms. The term $-ig\tilde{a}^\dagger\tilde{a}$ denotes the optical drive from stimulated Brillouin scattering, and the term $\sqrt{\xi}B_{\rm in}$ represents the microwave drive for the phonon field. The latter is determined by the extrinsic coupling rate $\xi$ (with unit [Hz]) and the length-normalized electromechanical drive amplitude $B_{\rm in}$ (with unit $[\sqrt{{\rm Hz/m}}]$).

The initial conditions of our system are set to be

$$\tilde{a}(z=0, t) = \tilde{a}_{\rm in}(t), \tag{S11}$$

where $\tilde{a}_{\rm in}$ is the incident optical envelope of our system. In the following text, we will discuss two separate situations in which either the optical drive or the microwave drive dominates the acoustic behaviors.

### A. Optical phase modulator

In Sec. IV, we introduce an optical phase modulator where the incident signal is a monochromatic light field with a frequency $\omega_0$ and an amplitude $a_0$, i.e. $\tilde{a}_{\rm in}(t) = a_0$. In this case, the phonon dynamics are primarily governed by the electromechanical drive, leading to the equation of motion for phonons

$$\frac{\partial\tilde{b}}{\partial t} + \chi_{\rm b}^{-1}(\Omega)\tilde{b} = \sqrt{\xi}B_{\rm in}, \tag{S12}$$

where we introduce the phonon susceptibility as $\chi_{\rm b}(\Omega) = \left[-i(\Omega - \Omega_0) + \Gamma/2\right]^{-1}$. In the steady-state approximation, the phonon amplitude is directly calculated as $b = \sqrt{\xi}\chi_{\rm b}B_{\rm in}$. Furthermore, using the input-output theory, we can obtain the electromechanical output of the acoustic drive, $B_{\rm out} = \sqrt{\xi}b = \xi\chi_{\rm b}B_{\rm in}$. Note that both $B_{\rm in}$ and $B_{\rm out}$ are directly transformed from or into microwave power input $P_{\rm RF}^{\rm in}$ and output $P_{\rm RF}^{\rm out}$ through the IDTs, with the transformation relationship expressed as

$$P_{\rm RF}^{\rm in} = \hbar\Omega L\,|B_{\rm in}(\Omega)|^2 / \eta_{\rm e}; \;\; P_{\rm RF}^{\rm out} = \hbar\Omega L\,|B_{\rm out}(\Omega)|^2 \cdot \eta_{\rm e}. \tag{S13}$$

Here, $L$ is the length of our electrically interfaced Brillouin-active waveguide. The microwave extinction, defined as $\eta_{\rm e} = 1 - |S_{11}(\Omega)|^2$, quantifies the efficiency of on-chip microwave coupling and is determined by the impedance design of the interdigital transducers (IDTs). In our experiments, we measure $|S_{11}(\Omega)|$ as the microwave reflection from a single IDT.

Correspondingly, the electromechanical coupling coefficient, which quantifies the power conversion efficiency from microwave to phonons, can be calculated as $\eta_{\rm em} = \sqrt{P_{\rm RF}^{\rm out}/P_{\rm RF}^{\rm in}}$. In our experiment, this coupling efficient is measured as the microwave transmission $|S_{21}|$ from one IDT to the other. When the microwave drive is on resonance, we obtain $\eta_{\rm em} = |S_{21}| = \eta_{\rm e}\cdot 2\xi/\Gamma$. The second factor of this parameter represents the on-chip microwave-to-acoustic conversion efficiency, and in our paper, we denote it as $\eta_{\rm m} = 2\xi/\Gamma$.

Next, we study the optical dynamics of this system. Plugging in the constant amplitude of the phonon field into eq. (S9), we arrive at

$$\left(\frac{\partial}{\partial t} + v_{\rm g}\frac{\partial}{\partial z}\right)\tilde{a} = -i2g|b|\cos\left(\Omega_0 t\right)\tilde{a}, \tag{S14}$$

By solving this equation, we can analyze the output light at position $z$

$$\tilde{a}(z, t) = a_0\exp\left\{\frac{i2g|b|}{\Omega_0}\left\{\sin\left[\Omega_0\left(t - \frac{z}{v_{\rm g}}\right)\right] - \sin\left(\Omega_0 t\right)\right\}\right\}. \tag{S15}$$

In our system, $\Omega_0 z/v_{\mathrm{g}} \ll 1$ given the short interaction length of our device ($z < 160 \ \mu\mathrm{m}$), the solution can be further simplified as a pure phase modulation

$$\tilde{a}(z,t) = a_0 \exp\left[-i\frac{2g|b|z}{v_{\mathrm{g}}}\cos\left(\Omega_0 t\right)\right].$$ (S16)

This is how we derived eq. (1) and obtain the value for modulation depth, $\beta = 2g|b|z/v_{\mathrm{g}}$, as presented in the main text. In our experiment, we measure and analyze the phase-modulated output light by decomposing it into a series of distinct optical tones, each with a frequency $\omega_n = \omega_0 + n\Omega_0$. The amplitude of $n$-th sideband at position $z$ can be obtained using Jacobi-Anger expansion, $e^{iz\cos\phi} = \sum_n i^n J_n(z)e^{in\phi}$. Hence we arrive at

$$\tilde{a}_n(z,t) = i^n a_0 J_n\left(-\frac{2g|b|z}{v_{\mathrm{g}}}\right)e^{-in\Omega_0 t}.$$ (S17)

Notice that the same dynamics can be derived using the cascading equations of motion for the traveling Brillouin scattering process [1, 5].

Combining eq.(S13) and (S17), we obtain the modulation depth on acoustic resonance

$$\beta = \sqrt{2\eta_{\mathrm{em}}G_{\mathrm{B}}P_{\mathrm{RF}}L}\sqrt{\frac{\omega_0}{\Omega_0}},$$ (S18)

where we introduce the Brillouin gain coefficient $G_{\mathrm{B}} = 4g^2/(v_{\mathrm{g}}^2\Gamma_0\hbar\omega_0)$ [1] to simplify the expression. When using a microwave input with a standard impedance of $R_0 = 50 \ \Omega$ to drive our system, the half-wave voltage $V_\pi$ of this phase modulation can be directly obtained from the modulation depth $\beta$ as

$$V_\pi = \pi\sqrt{\frac{R_0\Omega_0}{2\eta_{\mathrm{em}}G_{\mathrm{B}}L\omega_0}}.$$ (S19)

We can conclude that a higher electromechanical ($\propto \eta_{\mathrm{em}}$) or optomechanical ($\propto G_{\mathrm{B}}$) transduction rate, as well as a longer interaction length $L$, all contribute to a more efficient phase shifter.

When using this phase modulation for microwave-to-optical conversion, a small RF drive will be converted into $\pm 1$ sidebands in the system. The number of generated photons in one sideband, $n_{\mathrm{o}} = |a_{\pm 1}|^2 L$, can then be obtained by expanding the Bessel function to first order $J_1(x) \approx x/2$, which is given as

$$n_{\mathrm{o}} = \frac{1}{2}\eta_{\mathrm{em}}G_{\mathrm{B}}P_{\mathrm{RF}}L^2\frac{\omega_0}{\Omega_0}|a_0|^2 = \frac{1}{2}\eta_{\mathrm{em}}G_{\mathrm{B}}P_0L\frac{P_{\mathrm{RF}}L}{\hbar\Omega_0 v_{\mathrm{g}}},$$ (S20)

where we calculate the optical pump power as $P_0 = \hbar\omega_0 v_{\mathrm{g}}|a_0|^2$. The number of microwave photons $n_{\mathrm{RF}}$ that participate in this electo-optomechanical interaction can be calculated from the incident RF power $P_{\mathrm{RF}}$ and the time of flight $L/v_{\mathrm{g}}$, given as $n_{\mathrm{RF}} = (P_{\mathrm{RF}}L)/(\hbar\Omega v_{\mathrm{g}})$. Therefore the microwave-to-optical quantum conversion efficiency can then be obtained as

$$\eta = \frac{n_{\mathrm{o}}}{n_{\mathrm{RF}}} = \frac{1}{2}\eta_{\mathrm{em}}G_{\mathrm{B}}P_0L.$$ (S21)

### B. Optical-to-microwave transduction

This section focuses on the behavior of phonons driven solely by optical drives, as described by the equation

$$\frac{\partial\tilde{b}}{\partial t} + \chi_{\mathrm{b}}^{-1}(\Omega)\tilde{b} = -ig\tilde{a}^\dagger(z,t)\tilde{a}(z,t).$$ (S22)

In combination with eq. (S9), we arrive at the optical envelope solution [3]

$$\tilde{a}(z,t) = \tilde{a}_{\mathrm{in}}\left(t - \frac{z}{v_{\mathrm{g}}}\right)\exp\left\{-i2g\int \mathrm{d}z'\,\mathrm{Re}\left[\tilde{b}\left(z',t-\frac{z-z'}{v_{\mathrm{g}}}\right)\right]\right\},$$ (S23)

which features a purely real-valued integrand, implying that it introduces only a phase modulation. Consequently, $\tilde{a}^\dagger(z,t)\tilde{a}(z,t) = \tilde{a}_{\mathrm{in}}^\dagger\tilde{a}_{\mathrm{in}}$, indicating that the envelope remains constant as the system evolves. This result is in accordance with the conclusion derived from the Hamiltonian discussion. In the subsequent discussions, we utilize this result to analyze two distinct experiments described in the paper: the characterization of transduction efficiency and the measurement of microwave channelization.

### 1. Two-tone excitation

In the transduction efficiency characterization experiment, we inject two optical drive fields, each with the frequency $\omega_0$ ($\omega_{-1}$) and amplitude $a_0$ ($a_{-1}$) into the system. Considering the rotating wave approximation, the stationary solution of the phonon field is then obtained as

$$\tilde{b} = -ig\chi_{\mathrm{b}}(\Omega)a_0(0)^\dagger a_{-1}(0). \tag{S24}$$

Following the electromechanical transduction process described in eq. (S13), the microwave output on acoustic resonance is calculated as $P_{\mathrm{RF}}^{\mathrm{out}} = \frac{1}{2}\eta_{\mathrm{em}}G_{\mathrm{B}}P_0 L\hbar\Omega_0 v_{\mathrm{g}}|a_{-1}|^2$. Accordingly, the quantum efficiency of optical-to-microwave conversion is then

$$\eta = \frac{n_{\mathrm{RF}}}{n_{\mathrm{o}}} = \frac{1}{2}\eta_{\mathrm{em}}G_{\mathrm{B}}P_0 L, \tag{S25}$$

which is consistent with our result of the microwave-to-optical conversion, as expected.

### 2. Intensity-modulated light input

For the channelization experiment, we inject an intensity-modulated optical field into the system. We first consider the single-segment case, where the modulation signal is a single-frequency tone with the modulation frequency $\Omega$. Assuming the intensity modulator works at the quadrature point, the incident light can then be decomposed into a series of orders

$$a_{\mathrm{in}} = \sqrt{a_0(0)}\left(\frac{i}{\sqrt{2}} + \frac{1}{\sqrt{2}}\sum_n J_n(\beta_{\mathrm{IM}})e^{-in\Omega t}\right), \tag{S26}$$

where $\beta_{\mathrm{IM}}$ is the intensity modulation depth determined by the RF signal amplitude we dial into our commercial intensity modulator. Thus, the desired beatnote is directly produced from the $\pm 1$ sidebands, resulting in a local phonon field with strength $b_n = -g^*\chi_{\mathrm{b,n}}(\Omega)J_1(\beta_{\mathrm{IM}})|a_0(0)|^2$ at the $n$-th active region. The distinct microwave output is then given by

$$P_{\mathrm{RF,n}}^{\mathrm{out}}(\Omega) = \frac{\Omega}{\omega_0}\eta_{\mathrm{em,n}}G_{\mathrm{B,n}}J_1^2(\beta_{\mathrm{IM}})P_0^2 L\frac{(\Gamma_n/2)^2}{(\Omega - \Omega_n)^2 + (\Gamma_n/2)^2}, \tag{S27}$$

which shows adjustable acoustic responses and microwave readouts at distinct frequencies $\Omega_n$ with varying bandwidths $\Gamma_n$.

According to eq. (S23), the phase modulation stimulated by the acousto-optic interaction in the $n$-th active region is then given as

$$\tilde{a}(z,t) = \tilde{a}_{\mathrm{in}}\left(t - \frac{z}{v_{\mathrm{g}}}\right)\exp\left[-i\frac{2g|b_n|L}{v_{\mathrm{g}}}\cos(\Omega_n t)\right]. \tag{S28}$$

### 3. Microwave channelization

When the system features multiple active segments, the new Hamiltonian is described as

$$\mathcal{H} = \int \mathrm{d}z a^\dagger(z,t)\mathcal{L}_a a(z,t) + \sum_n \int_{z_{n-1}}^{z_n}\mathrm{d}z b_n^\dagger(z,t)\mathcal{L}_b b_n(z,t) + g\sum_n\int_{z_{n-1}}^{z_n}\mathrm{d}z a^\dagger(z,t)a(z,t)\left[b_n(z,t) + b_n^\dagger(z,t)\right], \tag{S29}$$

where $b_n$ is the phonon operator of the $n$-th region, ranging from $z_{n-1}$ to $z_n$, and $z_0 = 0$. For each segment, eq. (S22) and (S23) still hold, and the optical envelope output from each individual segment can be described as a function of the input to that segment, given as

$$\tilde{a}(z_n,t) = \tilde{a}_{n,\mathrm{in}}\left(t - \frac{z_n - z_{n-1}}{v_{\mathrm{g}}}\right)\exp\left\{-i2g\int_{z_{n-1}}^{z_n}\mathrm{d}z'\,\mathrm{Re}\left[\tilde{b}_n\left(z',t - \frac{z_n - z'}{v_{\mathrm{g}}}\right)\right]\right\}, \tag{S30}$$

TABLE S1. **The device parameters.** Notice that the source of the parameters is indicated in the footnote. Additionally, the columns with slashes are either not measured or not discussed in this work.

| Device parameters | Device 1 | Device 2 | Device 3 | Device 4 | Device 5 | Device 6 |
|---|---|---|---|---|---|---|
| The number of IDT teeth [a] | 7 | 5 | 3 | 3 | 3 | 3 |
| Membrane width $W$ (μm) [a] | 28.48 | 22.68 | 16.88 | 16.15 | 15.43 | 14.70 |
| $d$ (μm) [a] | 1.84 | 1.84 | 1.84 | 1.48 | 1.11 | 0.75 |
| $\Omega_0/(2\pi)$ (GHz) [b] | 3.644 | 3.642 | 3.655 | 3.746 | 3.654 | 3.631 |
| $\Gamma_0/(2\pi)$ (MHz) [b] | 12.65 | 10.24 | 7.39 | / | / | / |
| Acoustic $Q$-factor [b] | 288.01 | 355.66 | 494.59 | / | / | / |
| $\eta_{e,1} = 1 - |S_{11}(\Omega_0)|^2$ [b] | 2.92 % | 1.25 % | 0.78 % | / | / | / |
| $\eta_{e,2} = 1 - |S_{22}(\Omega_0)|^2$ [b] | 3.38 % | 2.12 % | 1.28 % | / | / | / |
| $\eta_{em} = |S_{12}(\Omega_0)|$ [b] | 0.0132 | $5.49\times10^{-3}$ | $2.69\times10^{-3}$ | / | / | / |
| $\eta_m$ [b c] | 0.419 | 0.337 | 0.270 | / | / | / |
| $g(\Omega_0)$ (Hz·$\sqrt{m}$) [d] | 1269.06 | 1972.31 | 1939.06 | / | / | / |
| $v_g$ (m/s) [d] | $8.33\times10^7$ | $8.33\times10^7$ | $8.33\times10^7$ | / | / | / |
| Simulated $G_B$ (W$^{-1}$·m$^{-1}$) [d] | 91.14 | 271.95 | 364.24 | / | / | / |
| Measured $G_B$ (W$^{-1}$·m$^{-1}$) [b e] | 27.45±1.60 | 66.84±3.63 | 50.64±1.47 | / | / | / |
| Maximum measured $\eta_{OE}$ [b f] | -59.79±0.25 | -59.24±0.24 | -63.49±0.13 | / | / | / |
| Maximum measured $\eta_{EO}$ [b g] | / | -54.16±0.24 | / | / | / | / |

[a] Device designs.
[b] Experimental measurements.
[c] Obtained as $\eta_m = \eta_{em}/\sqrt{\eta_{e,1}\eta_{e,2}}$.
[d] COMSOL simulations.
[e] Derived from the dual-pump optical-to-microwave transduction measurement.
[f] Measured at optical pump power of 16.12 dBm.
[g] Measured at optical pump power of 20.61 dBm.

where the input to $n$-th segment $\tilde{a}_{n,in}(t)$ is the output of the $(n-1)$-th segment, i.e. $\tilde{a}_{n,in}(t) = \tilde{a}(z = z_{n-1}, t)$. Hence the final output of the system can be obtained using the recursion method

$$\tilde{a}(z,t) = \tilde{a}_{in}\left(t - \frac{z}{v_g}\right)\exp\left\{-i2g\sum_n\int_{z_{n-1}}^{z_n}\mathrm{d}z'\,\mathrm{Re}\left[\tilde{b}_n\left(z', t - \frac{z_n - z'}{v_g}\right)\right]\right\},\tag{S31}$$

where each segment introduces a unique phase modulation determined by the acoustic properties of that segment, resulting in a constant optical envelope throughout the entire system. When all the active segments are of the same length $L$, we arrive at the output field

$$\tilde{a}(z,t) = \tilde{a}_{in}\left(t - \frac{z}{v_g}\right)\exp\left[-i\sum_n\beta_n\cos(\Omega_n t)\right],\tag{S32}$$

where $\beta_n = \frac{2g|b_n|L}{v_g}$ is the modulation depth of the $n$-th region.

## II. DESIGN IMPACTS ON DEVICE PERFORMANCE

In the main text, we present measurements from six devices with different designs: Device 1 for the characterization of electromechanical Brillouin scattering and optical phase modulation (Fig. 2,3), Device 2 for bidirectional optical-to-microwave transduction (Fig. 4), and device 3~6 for the microwave channelization (Device 4, 5, 6 were measured in Fig. 5c and Device 1, 2, 3 were measured in Fig. 5d). We present the device designs, as well as their corresponding simulation and measurement results, in Table S1. In this section, we provide detailed information about how these parameters are measured or simulated. We also investigate the impact of device designs on their performance and discuss how these studies can guide us in the future development of electrically interfaced Brillouin-active waveguides.

### A. Device dimensions

All the devices under study share the same basic structure, as illustrated in Fig. S1, where the 200-nm-thick aluminium IDT electrodes sit on a piezoelectric 480-nm-thick AlN film supported by a silicon membrane structure.

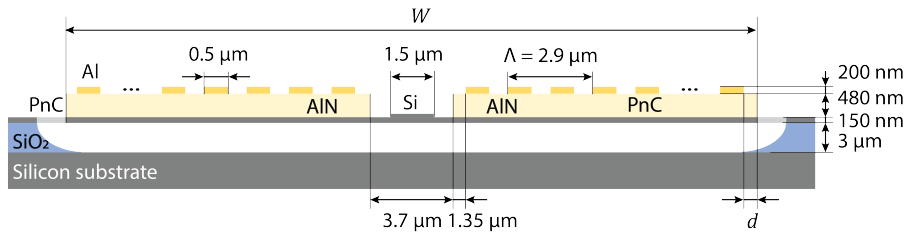

FIG. S1. **The basic device structure.**

To ensure both vertical and lateral acoustic confinement, the AlN/Si membrane is suspended and bordered laterally by a phononic crystal structure composed of circular holes with diameter (pitch) of 616 (800) nm to ensure lateral acoustic confinement. The IDT pitch, $\Lambda = 2.9$ μm, is carefully designed to optimize the electromechanical transduction of the target phonon at the frequency $\Omega_0$. The silicon optical waveguide has a width of 1.5μm. The AlN film is discontinued on top of the silicon optical waveguide in the center of the device by 3.7 μm to minimize the optical loss. The membrane width $W$, determined by the number of IDT teeth and the IDT to phononic crystal distance $d$, is used to adjust the acoustic resonance. In this paper, we studied one 7-tooth device (Device 1), one 5-tooth device (Device 2), and four 3-tooth devices (Device 3 to Device 6).

### B. Electromechanical transduction

The electromechanical transduction efficiency, $\eta_{em} = \eta_e \eta_m$, is determined by two factors, the microwave extinction $\eta_e$ and the target acoustic mode external coupling rate $\eta_m$. Here we compare and analyze $\eta_{em}$ across three devices with different number of IDT teeth, but the same sets of acoustic eigenmodes, and summarize the key considerations for improving electromechanical transduction when designing electrically interfaced Brillouin-active waveguides.

In out experiment, electromechanical transduction can be directly characterized by the microwave reflection and transmission of the two-IDT system, expressed as $\eta_e = 1 - |S_{11}|^2$ and $\eta_{em} = |S_{21}|$. Fig. S2 shows the four $S$-parameter measurements from Device 1, Device 2, and Device 3, which are three different devices with 7, 5, and 3 sets of IDT teeth, respectively, but with the same IDT to phononic crystal distance. The identical IDT to phononic crystal distance in these three devices results in these three devices featuring the same sets of high-$Q$ acoustic eigenmodes, namely A and B discussed in the main text. This is confirmed in our measurements, where all the $S$-parameters reveal two resonances at 3.53 GHz and 3.63 GHz. However, these three devices exhibit very different microwave extinction values, indicating varying capabilities of coupling microwave power into the system: the more IDT teeth a device has, the better the impedance matching to a standard 50 Ω input impedance of the RF probe used in the experiment, and the greater the efficiency of coupling microwave power into the system. Additionally, these three devices also display significantly different target acoustic mode external coupling rates, calculated as $\eta_m = \eta_{em}/\eta_e$. We find that the more IDT teeth a device has, the more well-defined the acoustic drive will be, and as a result, the better $\eta_m$ becomes. In summary, a higher number of IDT teeth in a device leads to more efficient electromechanical transduction.

It should also be noted that the microwave reflections from two IDTs within a single device are not completely identical, i.e., $|S_{11}| \neq |S_{22}|$. However, the variation is within 1%, suggesting that the two IDTs are relatively similar. This discrepancy can be attributed to several factors, including variations in probes, differences in fabrication between the two IDTs, and variations in contact between the two probe-IDT interfaces. To address this variation, we assume the microwave extinction as the geometric mean of the two IDT measurements, given as $\eta_e = \sqrt{\eta_{e,1}\eta_{e,2}}$ in our experiment.

### C. Brillouin gain

The Brillouin gain, $G_B = 4g^2/(v_g^2 \Gamma_0 \hbar \omega_0)$, is determined by two factors, the acousto-optic coupling rate $g$, and the acoustic loss rate $\Gamma$. By comparing and analyzing the results of the three devices presented in Fig. S3, we can also identify key considerations for improving the Brillouin gain when designing electrically interfaced Brillouin-active waveguides.

We begin by studying the acousto-optical coupling rate, $g$, of three devices using COMSOL simulations, which is

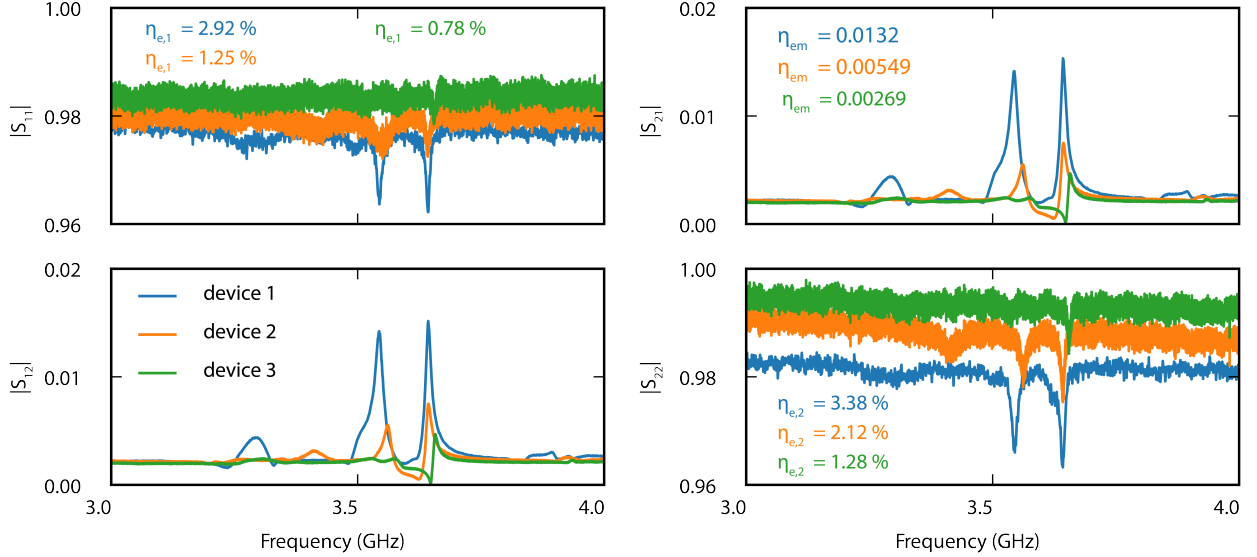

FIG. S2. **The electromechanical transduction evaluation.** The key parameters to determine electromechanical transduction, $\eta_{\rm em}$ and $\eta_{\rm e}$, can be directly derived from the microwave $S$-parameter measurements. Here we analyze and compare the performance from Device 1 (blue), Device 2 (orange), and Device 3 (green).

calculated as

$$g = \frac{1}{\epsilon_0} \sqrt{\frac{\omega_1}{2}} \sqrt{\frac{\omega_2}{2}} \sqrt{\frac{\hbar\Omega_0}{2}} \int d\mathbf{r}_\perp \left(D^i\left(\mathbf{r}_\perp\right)\right)^* D^j\left(\mathbf{r}_\perp\right) p^{ijkl}\left(\mathbf{r}_\perp\right) \frac{\partial u_0^k\left(\mathbf{r}_\perp\right)}{\partial r^l}, \tag{S33}$$

where $p^{ijkl}$ is the photoelastic tensor of silicon; $D$ is the electric displacement field of the optical mode, and $u$ is the displacement field of acoustic mode driven by the IDT. The simulation result is presented in Fig. S3a, with the electromechanical resonant point corresponding to mode A marked in circles. We notice that the acousto-optic overlap for device 1 (blue) is much lower than device 2 (orange) and device 3 (green), due to the dilution of the acoustic mode energy over a much larger membrane. We have also observed a mismatch between the frequency for the optimum acousto-optic coupling rate and the electromechanical resonant mode A for Device 1, whereas Devices 2 and 3 are well aligned. This results in Devices 2 and 3 having a significantly larger Brillouin gain compared to Device 1.

The acoustic $Q$-factor is another factor that affects the Brillouin gain and is proportional to it. As shown in Fig. 5d, increasing the number of metal teeth in the acoustic membrane significantly increases the acoustic loss, which in turn deteriorates the acoustic $Q$-factor. To determine the impact of these factors on the Brillouin gain for each device, we combined the simulated values of $g$ with the measured acoustic loss rate, $\Gamma$, to derive theoretical predictions for the Brillouin gain. Our calculations yielded a simulated $G_{\rm B}$ of 91.14 W$^{-1}\cdot$m$^{-1}$ for Device 1, 271.95 W$^{-1}\cdot$m$^{-1}$ for Device 2, and 364.24 W$^{-1}\cdot$m$^{-1}$ for Device 3.

We verify our simulations with the dual-pump setup to measure stimulated Brillouin processes. As illustrated in Fig. 4a, both optical drives are set to the same power level, $P_0 = P_{-1}$. In this experiment, the transduced microwave power is directly proportional to the square of optical input power, as given by $P_{\rm RF}^{\rm out}/P_0^2 = \eta_{\rm em}G_{\rm B}L\Omega_0/(2\omega_0)$. Fig. S3b presents our measurements of transduced microwave power across three devices as a function of optical input power $P_0$. Among the three devices, Device 3 (green) exhibits the lowest optical-to-microwave conversion rate due to its unsatisfactory electromechanical transduction efficiency. In contrast, Devices 1 (blue) and 2 (green) shows comparable performance, with Device 1 having higher electromechanical transduction but lower Brillouin gain, and Device 2 having lower electromechanical transduction but higher Brillouin gain. We extracted the actual Brillouin gain $G_{\rm B}$ values from the linear fits with a slope equal to 2 (black). These values are 27.45 W$^{-1}\cdot$m$^{-1}$ for Device 1, 66.84 W$^{-1}\cdot$m$^{-1}$ for Device 2, and 50.64 W$^{-1}\cdot$m$^{-1}$ for Device 3. The extracted Brillouin gain values were at least a factor of 4 smaller than our simulated values, indicating significant potential for improving the design of future electrically interfaced Brillouin-active waveguides. Several factors contribute to the discrepancy in the extracted gain values, including misalignment between the optimal electromechanical resonance and acoustic modes, which leads to suboptimal acousto-optical overlaps, fabrication imperfections resulting in imperfect overlaps, and the reduced photoelastic constants of strained silicon.

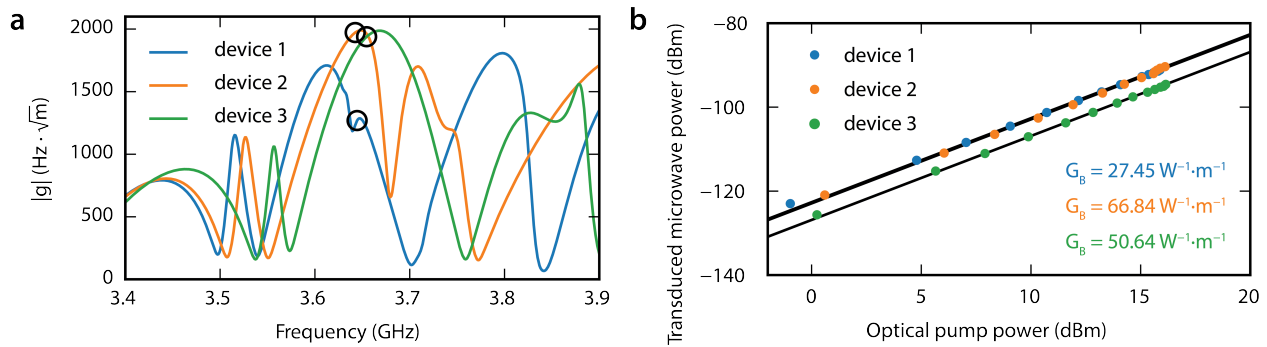

FIG. S3. **The acousto-optical coupling evaluation. a,** Simulated acousto-optical coupling rate $g$ as a function of frequency for Device 1, Device 2, and Device 3. The electromechanical resonance, mode A, is marked in black circle for each of the device. **b,** The optical-to-microwave conversion effciency meaurements across three devices. The meaurements are conducted through a dual-pump experimental setup as illustrated in Fig. 4a.

### D. The interaction length

The selection of the interaction length, $L$, involves balancing the tradeoff between the higher conversion efficiency associated with longer interaction lengths and the benefits of better phase matching and lower optical propagation loss associated with shorter interaction lengths. In this study, we have chosen an interaction length of $L = 160$ μm for all investigated devices. However, this interaction length can be extended by at least 10 to 100 times without incurring significant losses due to optical propagation or acousto-optic phase-mismatch. Such an extension could lead to a 10-20 dB enhancement in quantum efficiency for bidirectional optical-to-microwave conversion or up to 100 microwave output channels in microwave channelization.

The optical loss of the system was evaluated based on our previous measurements in [6], which found the optical mode propagation loss to be $\alpha = 7.00$ m$^{-1}$. This value translates to a negligible optical propagation loss of -4.86×10$^{-3}$ dB for each active segment, suggesting that achieving 100 channels in a linear device would still result in a acceptable overall loss of -0.486 dB.

The phase-mismatch of the device was also evaluated. According to the phase-matching diagram in Fig. 1d, the wavevector-mismatch was calculated to be $\Delta q = \Omega_0/v_g = 43.75$ m/s. This suggests a negligible phase-mismatch loss of sinc$^2(\Delta q L/2) \sim$ -1.75 ×10$^{-4}$ dB for each active region, indicating that attaining 100 channels in a linear device would still yield a reasonable overall loss of -1.83 dB.

## III. RESONANT ENHANCEMENT OF ELECTROMECHANICAL TRANSDUCTION

Our device demonstrates significant electro-mechanical and acousto-optic responses, despite the large impedance mismatch from the standard input impedance of 50 Ω and the presence of less than 10 IDT teeth. This significant enhancement is achieved through the acoustic resonance that we designed into the system. In this section, we study and quantify the electromechanical transduction enhancement resulting from the acoustic resonance by analyzing the admittance of the electromechanical transducer.

The admittance is obtained from the microwave $S_{11}$ measurement of Device 1. According to the Butterworth–Van Dyke model [7], the total electrical admittance of the transducer is given as

$$Y = i\Omega C_s + G(\Omega) + iB(\Omega).$$ (S34)

Here $C_s$ is the static capacitance, $G(\Omega)$ and $B(\Omega)$ are the conductance and the susceptance as a function of RF frequency $\Omega$. For a transducer with $N$ finger-pairs, the static capacitance is given as:

$$C_s = Nw_T(\epsilon_0 + \epsilon_p),$$ (S35)

where $w_T$ is the transducer width (aperture) and $\epsilon_p$ is the effective permittivity of the piezoelectric material. In our system, $N = 3.5$, $w_T = 160$ μm, and the AlN permittivity $\epsilon_p \approx 9\epsilon_0$, which leads to the static cpacitance of 0.050 pF. This agrees with the off-resonant fitting of the imaginary admittance (the black line in Fig. S4a), which gives $C_s = 0.036$ pF.

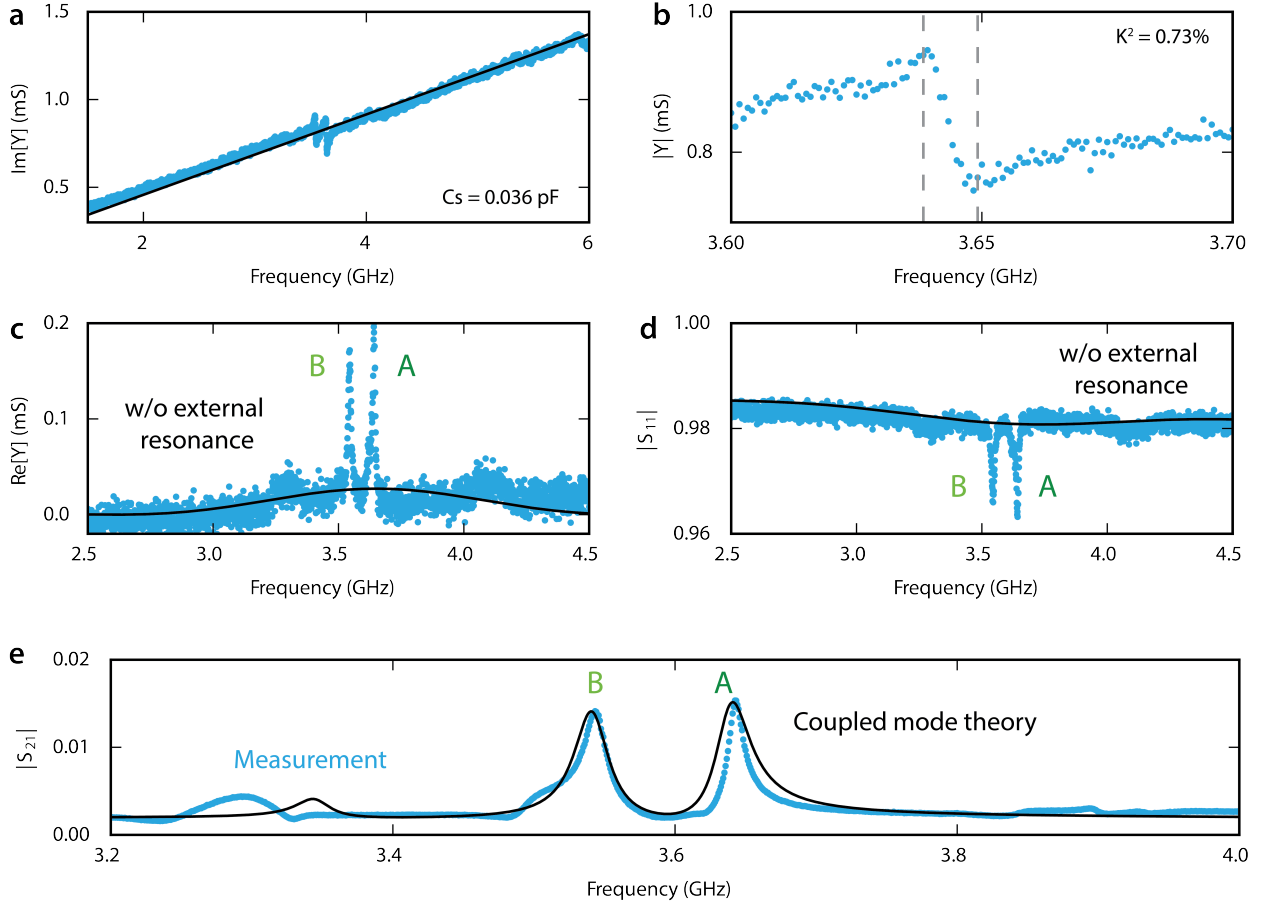

FIG. S4. **The electrical properties of IDT 1 in device 1.** **a**, The imaginary part of the transducer admittance reveals a static capacitance of $C_s = 0.036$ pF, which aligns well with our theoretical calculation of 0.0050 pF. **b**, The absolute value of transducer admittance exhibits a $k_{eff}^2$ of 0.73%, in agreement with the COMSOL simulated value of 1.03%. **c**, We compared the real part of the transducer admittance obtained from measurement (blue) with the theoretical calculation from BVD model (black). The difference between these two curves reveals the resonant enhancement resulting from the acoustic resonances in the device. **d**, We compared the RF reflection $|S_{11}|$ from measurement (blue) with the theoretical calculation from BVD model (black). **e**, The microwave transmission $|S_{21}|$ obtained from measurement (blue) and the compled mode theory (black).

The radiation conductance $G(\Omega)$ and the motional susceptance $B(\Omega)$ are predicted by the reflection-free impulse response model (IRM) [8] as

$$G(\Omega) = G_0 \text{sinc}^2 x, \ B(\Omega) = G_0 \left( \frac{\sin 2x - 2x}{2x^2} \right).$$ (S36)

where $x \equiv N\pi \frac{\Omega - \Omega_0}{\Omega_0}$. The peak conductance is determined by $G_0 = 8\Omega_0/(2\pi)k_{eff}^2 C_s N$, where $k_{eff}^2$ is the electromechanical coupling coefficient. For a piezoelectric resonator with resonance frequency $\Omega_r$ and antiresonance frequency $\Omega_a$ (marked as the dashed lines in Fig. S4b), the electromechanical coupling coefficient is calculated as [7]

$$k_{eff}^2 = \frac{\pi^2}{4} \left( 1 - \frac{\Omega_r}{\Omega_a} \right).$$ (S37)

According to our measurement in Fig. S4b, $k_{eff}^2 = 0.73\%$. We implemented similar calculation in COMSOL Multiphysics through the acoustic eigenmode simulation using open (free) and short (metallized) boundary conditions, which consistently gives $k_{eff}^2 = 1.03\%$.

Based on the static cpacitance $C_s$ and the electromechanical coupling coefficient $k_{eff}^2$ that we obtained from Fig. S4a and b, the real part of the admittance is calculated using eq. (S34,S36) as the black line in Fig. S4c. The presence of multiple external acoustic resonances significantly increases the 'effective conductance' when the transducer is

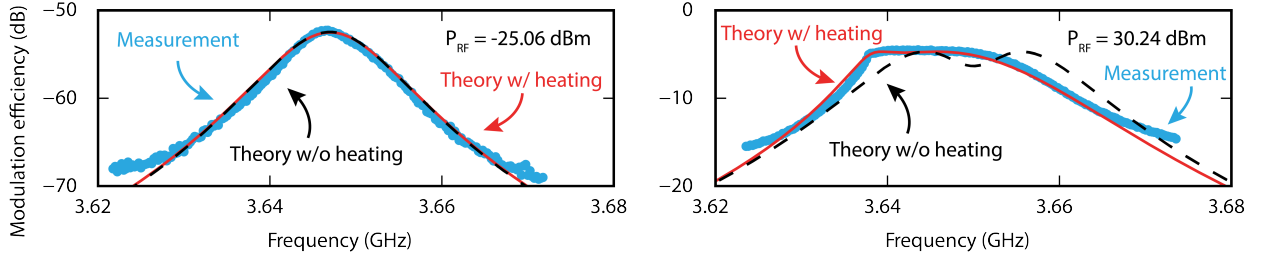

FIG. S5. **The thermal shifting of acoustic resonance under large microwave drive.** We introduce two theoretical models to analyze the thermal behavior of the system: the black dashed line represents a model without considering frequency shifts, and the red solid line shows a the impedance of eq. (S40) that describes thermal shifting. At low RF drive (left panel), the two models yield identical results and agree well with the measurements (blue), indicating negligible thermal heating. However, at high RF drive (right panel), the resonant frequency is red-shifted, resulting in a flattened spectrum. From the fitting, we extract a thermal heating coefficient of -260 MHz/W, which implies a maximum frequency shift of 8.27 MHz while sweeping frequency at this microwave input power. The system exhibits significant deviation from the ideal model, represented by the black solid line, with the absence of first sideband saturation.

operated on resonance, as denoted by the blue data points in Fig. S4c. For resonance A that we highlighted in Fig. 2, the conductance is improved by a factor of 7 compared with the BVD model without external resonances, resulting in a much better match to the impedance of the standard 50 Ω RF probe. Using the corresponding admittance to calculate the microwave reflection coefficient $|S_{11}|$ and comparing it to the measurement in Fig. S4d, we found that the microwave extinction $\eta_e$ has improved by approximately four times due to the acoustic resonance.

We developed a simplified model using coupled mode theory to explain the resonant enhancement resulting from the acoustic resonances observed in the system, as depicted in Fig. S4e. In our model, we assume a series of high-$Q$ mechanical modes are present in our system, with each mode separated by the FSR determined by the sound velocity and acoustic membrane width. The inclusion of metal gratings and multiple acoustic reflectors in the device leads to undesired coupling between adjacent high-$Q$ modes, resulting in uneven microwave transmission and varying acoustic linewidths for each distinct acoustic mode. Our theoretical predictions (black) align well with the experimental measurements (blue), indicating that the acoustic mode frequency can be effectively controlled through the acoustic membrane design. This can be achieved by adjusting the acoustic membrane width to alter the signal FSR or by designing the reflectors' position to regulate the intermodal coupling.

## IV. THERMAL BISTABILITY UNDER LARGE MICROWAVE DRIVE

In our phase modulator measurement, we attribute the spectral distortion observed in Fig. 3a to the thermal shifting of phonons. In this section, we offer a comprehensive analysis of this thermal shifting, presenting evidence from experimental results and theoretical derivations.

To solve this problem, we make several assumptions. First, we assume that all the on-chip microwave power that is not converted into acoustic power is instead absorbed as heat, which is determined as

$$P_{\text{abs}}(\Omega) = \eta_e P_{\text{RF}}^{\text{in}} - P_a = \left(1 - \frac{\xi}{(\Omega - \Omega_0 - \delta\Omega)^2 + \frac{\Gamma^2}{4}}\right) \eta_e(\Omega) P_{\text{RF}}^{\text{in}},$$

(S38)

where $\delta\Omega$ is the frequency shift of the acoustic resonance. It is important to note that the microwave extinction $\eta_e(\Omega)$ is a function of the incident microwave frequency and should also incorporate the thermal shift of the acoustic resonance as

$$\eta_e(\Omega) = \eta_e(\Omega_0) \frac{\frac{\Gamma^2}{4}}{(\Omega - \Omega_0 - \delta\Omega)^2 + \frac{\Gamma^2}{4}}.$$

(S39)

Secondly, we assume that, to the first order, the frequency shift of the acoustic resonance is proportional to the absorbed power, written as $\delta\Omega = c_{\text{heat}} P_{\text{abs}}(\Omega)$, where $c_{\text{heat}}$ is the heating coefficient with unit Hz/W. As a result, the absorption power $P_{\text{abs}}$ can be solved as an expression that is dependent on the detuned frequency $\Omega - \Omega_0$ and

the heating coefficient $c_{\text{heat}}$. If we plug the expression of the $P_{\text{abs}}$ back into the phonon amplitude, the modulation efficiency of the 1st sideband can then be expressed by the absorption power $P_{\text{abs}}$ as

$$\frac{|a_1|^2}{|a_0|^2} = \left| J_1 \left( \frac{2gL}{v_{\text{g}}} \sqrt{ \frac{\xi \eta_{\text{e}}(\Omega)}{(\Omega - \Omega_0 - c_{\text{heat}} P_{\text{abs}})^2 + \frac{\Gamma^2}{4}} \frac{P_{\text{RF}}^{\text{in}}}{\hbar \Omega L} } \right) \right|^2. \tag{S40}$$

In Fig.S5, we present our theoretical calculations alongside the measurements of the first sideband modulation spectra. We introduce two theoretical models: the dashed black line represents the model without frequency shifting from thermal heating, and the solid red line corresponds to the theoretical fitting derived in eq. (S40), accounting for frequency shifts due to thermal heating. We observe that at low microwave drive ($P_{\text{RF}}^{\text{in}} = $ -25.06 dBm), where thermal shifting is negligible, the two theoretical models converge into an identical trace that agrees well with the experimental measurements. However, at high microwave drive power ($P_{\text{RF}}^{\text{in}} = 30.24$ dBm), where thermal shifting becomes dominant in the experiment, we observe a significant red-shifting of the resonant frequency of the phonon modes in the red line, which is in good agreement with the flattened measured spectra. According to our fitting, we extract a linewidth broadening to 20 MHz and a thermal shifting parameter $c_{\text{heat}}$ of 260 MHz/W. This indicates that at this microwave drive input, the maximum thermal absorption while sweeping the frequency is about 31.8 mW, which corresponds to a frequency shift of 8.27 MHz. Our model effectively explains why our system does not exhibit phase modulation saturation at high microwave power, as highlighted by the contrast between the resonant dip in the ideal model without thermal heating (dashed line) and the actual system. At present, our thermal model does not adequately fit the higher-order sideband modulation spectra. We hypothesize that this may be due to the intricate behavior exhibited by the higher-order sidebands, resulting from a combination of effects, such as IDT overtones, the presence of unwanted acoustic modes in the system, and the heating effect. However, a more comprehensive understanding of this issue necessitates further investigation.


\let\clearpage\relax

\end{document}